%% file: main.tex
\PassOptionsToPackage{hyphens}{url} %
\documentclass[manuscript]{acmart}
\setcopyright{none}
\renewcommand\footnotetextcopyrightpermission[1]{}
\acmJournal{PACMHCI}
\acmVolume{0}
\acmNumber{CSCW}
\acmArticle{0}
\acmMonth{0}
\acmYear{2027}
\usepackage{graphicx}
\usepackage{booktabs}
\usepackage{pifont} %
\usepackage{amsmath,amssymb}
\usepackage{xspace}
\usepackage{soul}
\usepackage{cleveref}
\Crefname{section}{Section}{Sections}
\Crefname{subsection}{Section}{Sections}
\Crefname{table}{Table}{Tables}
\Crefname{figure}{Figure}{Figures}
\Crefname{appendix}{Appendix}{Appendices}
\Crefname{subappendix}{Appendix}{Appendices}
\Crefname{subsubappendix}{Appendix}{Appendices}
\definecolor{figgray}{HTML}{B0B0B0}
\definecolor{figblue}{HTML}{5B8DB8}
\definecolor{figorange}{HTML}{C1440E}
\definecolor{figdark}{HTML}{333333}
\newcommand{\swatch}[1]{\textcolor{#1}{\rule{1.1ex}{1.1ex}}}
\newcommand{\mcirc}[1]{\textcolor{#1}{\raisebox{0.18ex}{\scalebox{1.15}{$\bullet$}}}}
\newcommand{\msq}[1]{\textcolor{#1}{\scalebox{0.8}{$\blacksquare$}}}
\newcommand{\mdia}[1]{\textcolor{#1}{\scalebox{0.9}{$\blacklozenge$}}}

\newcommand{\ours}{the distribution-trained model\xspace}
\newcommand{\Ours}{The distribution-trained model\xspace}

\makeatletter
\definecolor{parabg}{gray}{0.88}
\sethlcolor{parabg}
\newcommand{\BfPara}[2][]{\par\vspace{1ex}\noindent\textbf{\hl{#2.}}%
  \if\relax\detokenize{#1}\relax\else\phantomsection\label{#1}%
  \immediate\write\@auxout{\string\newlabel{pt@#1}{{#2}{\thepage}{}{}{}}}\fi\ }
\makeatother
\newcommand{\Pref}[1]{\Cref{#1}, ``\ref{pt@#1}''}
\newcommand{\Tref}[1]{``\ref{pt@#1}''}
\newcommand{\coderepo}{\url{https://github.com/arslanbisharat/severity}}
\usepackage{tikz}
\usetikzlibrary{arrows.meta,positioning,fit,backgrounds,shapes.geometric}
\definecolor{cmild}{HTML}{5B8DB8}
\definecolor{cmod}{HTML}{E3A21A}
\definecolor{csev}{HTML}{C1440E}
\definecolor{cgray}{HTML}{B0B0B0}
\newcommand{\person}[4][1]{%
  \begin{scope}[shift={(#2,#3)},scale=#1]
    \fill[#4] (0,0.48) circle (0.17);
    \fill[#4] (-0.3,0) arc (180:0:0.3) -- cycle;
  \end{scope}}
\newcommand{\mo}[1]{}\newcommand{\arslan}[1]{}\newcommand{\manny}[1]{}\newcommand{\yas}[1]{}
\newcommand{\hall}[1]{}\newcommand{\khan}[1]{}\newcommand{\kate}[1]{}\newcommand{\mujtaba}[1]{}

\begin{document}

\title{Calibrate, Don't Retrain: Decomposing the Value of Annotator Disagreement in Cyberbullying Severity}

\author{Arslan Bisharat}
\affiliation{\institution{Loyola University Chicago}\country{USA}}
\email{marslan@luc.edu}
\author{Katelyn Skees}
\affiliation{\institution{Arizona State University}\country{USA}}
\email{kskees@asu.edu}
\author{Manuel Sandoval Madrigal}
\affiliation{\institution{Loyola University Chicago}\country{USA}}
\email{msandovalmadrigal@luc.edu}
\author{Mujtaba Nazari}
\affiliation{\institution{Loyola University Chicago}\country{USA}}
\email{mnazari@luc.edu}
\author{Ayaan Khan}
\affiliation{\institution{Loyola University Chicago}\country{USA}}
\email{akhan95@luc.edu}
\author{Deborah L. Hall}
\affiliation{\institution{Arizona State University}\country{USA}}
\email{d.hall@asu.edu}
\author{Mohammed Abuhamad}
\affiliation{\institution{Loyola University Chicago}\country{USA}}
\email{mabuhamad@luc.edu}
\author{Yasin N. Silva}
\affiliation{\institution{Loyola University Chicago}\country{USA}}
\email{ysilva1@luc.edu}

\begin{abstract}
Machine learning models are often trained on labeled data produced by multiple human annotators. When those annotators disagree, their labels are commonly collapsed into a single majority vote, which discards information about the underlying disagreement. We examine the cost of this practice for cyberbullying severity, a subjective judgment on which annotators disagree heavily. In the corpus we study, all 5 annotators assigned the same severity label to only 4.1\% of comments. Turning 5 judgments into one label hides this disagreement and obscures how much the annotators differed in their assessments. In this paper, we ask whether a model can preserve this information by predicting the distribution of annotator judgments instead of a single majority label. We find 3 main results. First, accuracy against the majority label cannot tell competent models apart: 2 models with similar accuracy can differ greatly in how well they match the distribution of annotator judgments. Second, predicting the full distribution requires little extra effort. Calibrating an existing model, which adjusts its output probabilities after training, captures most of the improvement without retraining, across 3 datasets covering severity, abuse, and irony, while training on label distributions adds a small further gain. Third, we examine a dataset that includes annotator demographics, where a single shared model has one advantage and one limit. The advantage is that a model predicting the distribution of judgments matches each demographic group's judgments significantly better than a model trained on majority labels, so both groups gain. The limit is that it does not narrow the gap in performance between the groups, because one shared model cannot encode which group is right on a given comment. Preserving annotator disagreement is therefore a low-cost step toward fairer moderation, though not a complete one.
\end{abstract}


\maketitle

\input{sections/intro}
\input{sections/background}
\input{sections/methods}
\input{sections/results}
\input{sections/discussion}

\bibliographystyle{ACM-Reference-Format}
\bibliography{references}

\appendix
\input{sections/appendix}

\end{document}

%% file: sections/intro.tex
\section{Introduction}
\label{sec:intro}

Labeled datasets are essential for training machine learning models, and many are labeled by several human annotators whose judgments are then combined into one label per item \citep{aroyo2015truth, uma2021learning}. For cyberbullying, models trained on such data grade how severe a comment is, because a moderator deciding what to do about a comment needs to know how harmful it is \citep{jurgens2019just, schoenebeck2023online}. A mild insult may call for a different response than a credible threat. The moderator then sees a single severity label and acts on it.
That single label hides disagreement among human annotators. In the corpus we study, 5 annotators labeled each comment and all 5 chose the same severity label for only 4.1\% of comments, so for the other 95.9\% at least 1 annotator disagrees with the rest. The standard process reduces the 5 labels/levels into a single majority label. For example, a comment labeled \textit{``moderate''} by 3 annotators and \textit{``severe''} by 2 is treated the same as a comment labeled \textit{``moderate''} by all 5 annotators. %
Severity is a subjective judgment that depends on how annotators interpret context, relationships, and intent. These interpretations can differ among annotators \citep{sap2022annotators, davani2022dealing, haimson2021disproportionate}. This disagreement can provide potentially useful information about how people perceive the comment differently. Reducing their labels to a majority label discards this information and forces the model to learn a single consensus label, so the range of human judgments is lost \citep{jurgens2019just, scheuerman2021framework, pavlick2019inherent, plank2022problem}. In this work, we examine preserving annotator disagreement when building models. 
We represent disagreement as the distribution of labels across severity levels. For example, if 3 annotators choose \textit{``moderate''} and 2 choose \textit{``severe''}, the target distribution is 0\% \textit{``mild''}, 60\% \textit{``moderate''}, and 40\% \textit{``severe''}. 
The model uses the distribution instead of a single majority label (\Cref{fig:overview}).

\input{figures/fig_overview}

Prior work has argued that majority labels can hide meaningful disagreement and that models should better represent the range of human judgments \citep{baan2022stop, plank2022problem}, and methods exist for learning from annotators' labels \citep{uma2021learning, davani2022dealing}. 
However, there are open challenges and questions that we address in this work.
First, %
 \textit{when annotators disagree, can accuracy metrics indicate a robust model when trained on the majority label?} 
 Second, \textit{can an existing model report the distribution of annotator labels without being retrained, and how much does retraining add?} 
Third, \textit{does reporting that distribution represent each demographic group's judgments more fairly?}
To answer these questions, we build 11 systems that vary architecture and supervision one choice at a time \citep{cao2020coral, shi2023corn, fornaciari2021beyond}. 
We evaluate each system using both majority-label accuracy and its agreement with the full distribution of human labels.

Our findings show the following.\ding{182} Accuracy against the majority label cannot indicate competent models as two models with the same accuracy can differ greatly in matching the distribution of annotator labels. %
\ding{183} An existing model can report distribution-based metric without retraining. With a small calibration step, a model learns the label distributions with the same performance as a fully distribution-aware retrained model, which adds only a small further gain on 3 datasets covering \textit{severity}, \textit{abuse}, and \textit{irony}.
\ding{184} Reporting the distribution helps every group and favors none. On a dataset with annotator demographics, a model that reports the label distribution matches both men's and women's judgments better than a majority-trained model, while the gap between the groups stays the same. The gap stays because 1 shared model gives 1 distribution for each comment, so it cannot give men's and women's judgments of the same comment different weight.
These results show that preserving annotator disagreement is both useful and inexpensive. It provides a better way to evaluate models when human judgments are divided and it can be added to an existing model through calibration without retraining.

%% file: figures/fig_overview.tex
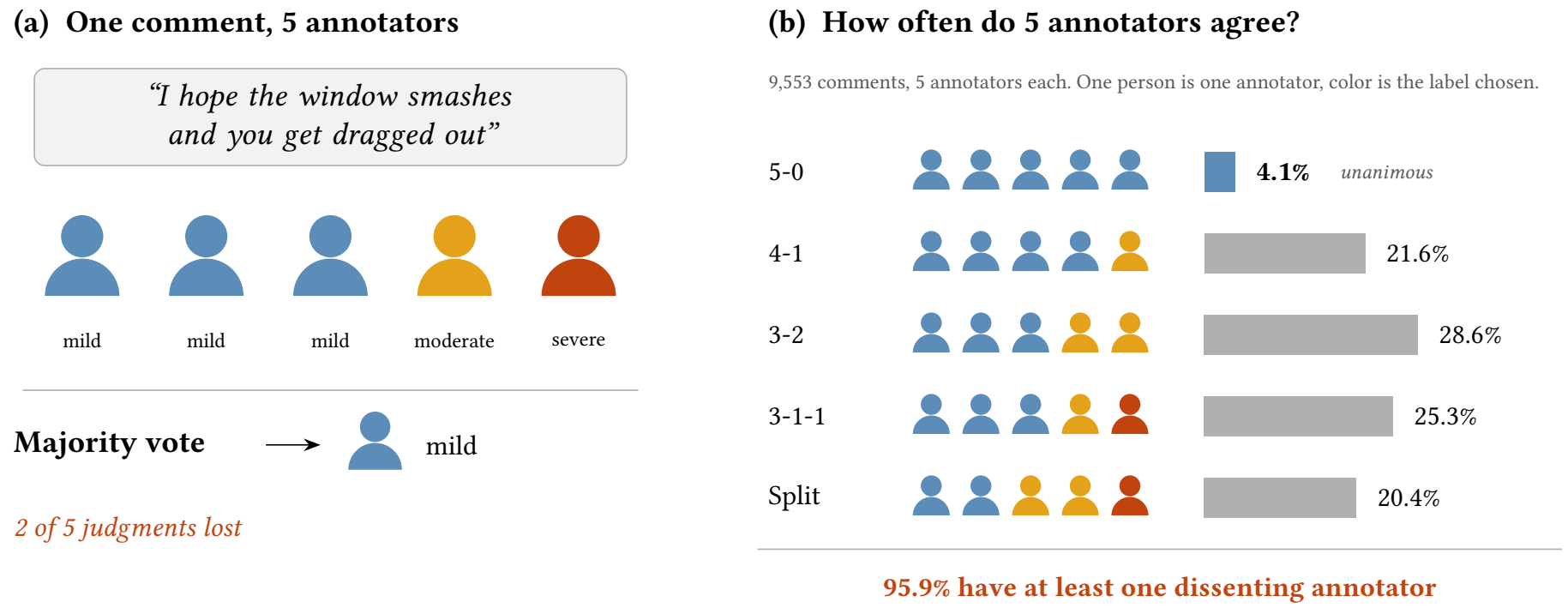
\begin{figure}[!htbp]
\centering
\resizebox{0.8\textwidth}{!}{%
\begin{tikzpicture}[font=\small]
\node[anchor=north west,font=\bfseries] at (0,0) {(a)\; One comment, 5 annotators};
\node[draw=cgray,fill=black!5,rounded corners=3pt,inner xsep=8pt,inner ysep=4pt,font=\itshape,text width=5.4cm,align=center] at (3.3,-1.2) {``I hope the window smashes and you get dragged out''};
\person[1.25]{0.8}{-3.0}{cmild}   \node[font=\scriptsize] at (0.8,-3.45) {mild};
\person[1.25]{2.05}{-3.0}{cmild}  \node[font=\scriptsize] at (2.05,-3.45) {mild};
\person[1.25]{3.3}{-3.0}{cmild}   \node[font=\scriptsize] at (3.3,-3.45) {mild};
\person[1.25]{4.55}{-3.0}{cmod}   \node[font=\scriptsize] at (4.55,-3.45) {moderate};
\person[1.25]{5.8}{-3.0}{csev}    \node[font=\scriptsize] at (5.8,-3.45) {severe};
\draw[cgray] (0.2,-3.95) -- (6.4,-3.95);
\node[anchor=west,font=\bfseries] at (0,-4.5) {Majority vote};
\draw[-{Stealth}] (2.65,-4.5) -- (3.2,-4.5);
\person[0.9]{3.75}{-4.75}{cmild}
\node[anchor=west] at (4.15,-4.5) {mild};
\node[anchor=west,csev,font=\itshape\small] at (0,-5.35) {2 of 5 judgments lost};
\node[anchor=north west,font=\bfseries] at (7.6,0) {(b)\; How often do 5 annotators agree?};
\node[anchor=north west,font=\scriptsize,text=black!65] at (7.6,-0.65) {9,553 comments, 5 annotators each. One person is one annotator, color is the label chosen.};
\foreach \k/\pat/\p/\pct/\w in {0/{5-0}/{cmild,cmild,cmild,cmild,cmild}/4.1/0.31,
                               1/{4-1}/{cmild,cmild,cmild,cmild,cmod}/21.6/1.62,
                               2/{3-2}/{cmild,cmild,cmild,cmod,cmod}/28.6/2.15,
                               3/{3-1-1}/{cmild,cmild,cmild,cmod,csev}/25.3/1.90,
                               4/{Split}/{cmild,cmild,cmod,cmod,csev}/20.4/1.53} {%
  \pgfmathsetmacro{\yy}{-1.75-0.82*\k}
  \node[anchor=west] at (7.6,\yy) {\pat};
  \foreach \c [count=\j from 0] in \p { \person[0.62]{9.35+0.5*\j}{\yy-0.18}{\c} }
  \ifnum\k=0 \fill[cmild] (12.1,\yy-0.2) rectangle ++(\w,0.4); \else \fill[cgray] (12.1,\yy-0.2) rectangle ++(\w,0.4); \fi
  \node[anchor=west] at (12.2+\w,\yy) {\ifnum\k=0 \textbf{\pct\%}\else\pct\%\fi};
}
\node[anchor=west,font=\scriptsize\itshape,text=black!65] at (13.35,-1.75) {unanimous};
\draw[cgray] (7.6,-5.55) -- (15.7,-5.55);
\node[csev,font=\bfseries\small] at (11.65,-5.95) {95.9\% have at least one dissenting annotator};
\end{tikzpicture}
}
\caption{Why a majority vote loses information. (a) A comment labeled mild by 3 annotators, moderate by 1 and severe by 1 becomes the single label mild, and the other 2 judgments are lost. (b) The share of the 9,553 comments in each agreement pattern. Only 4.1\% are unanimous, so 95.9\% have at least one annotator who disagrees with the rest.}
\label{fig:overview}
\end{figure}

%% file: sections/background.tex
\section{Background}
\label{sec:background}

This section reviews what is known about cyberbullying severity, why annotators disagree about it, how models can learn from that disagreement, and how it affects moderation and calibration. It ends with the 3 problems that our questions address.

\BfPara{Cyberbullying detection}
Computational work on cyberbullying and abusive language has concentrated on detection, the binary question of whether a message is harmful or not \citep{dinakar2011modeling, hosseinmardi2015detection, vanhee2018automatic}. Reviews identify inconsistent taxonomies and poor inter-annotator agreement as central bottlenecks \citep{rosa2019automatic, vidgen2020directions}, and crowdsourcing studies find that abusive content routinely receives split annotations \citep{founta2018large}. Previous work has also found that context-aware detectors, which analyze a comment along with its surrounding thread, improve cyberbullying detection on Instagram comment threads \citep{bisharat2026spectrumnet}.

\BfPara{Severity beyond detection}
Several lines of work call for moving beyond binary detection, which only labels a message as cyberbullying or not, since moderation needs severity-aware tools \citep{jurgens2019just}, severity is contextually negotiated \citep{scheuerman2021framework}, and victims of harassment ask for responses graduated by severity \citep{schoenebeck2023online}. The few systems that grade severity treat it as classification against a single majority label, so it remains open whether one label can represent the severity judgments of the people who annotate a comment.

\BfPara{Severity is contested in human judgment}
Research outside of computer science gives reason to expect disagreement about severity. The perception of harassment differs across annotator groups, including by gender \citep{rotundo2001meta} and age (for example, adolescents disagree about which behaviors count as cyberbullying and how serious they are \citep{menesini2012cyberbullying}), and the harm to victims depends on context and repetition and not on wording alone \citep{kowalski2014bullying}. Cyberbullying research also stresses that thread participants occupy distinct roles such as bully, victim, defender, and bystander \citep{salmivalli2010bullying}. Within NLP, annotator identity shapes hate speech labels \citep{waseem2016racist, sap2022annotators}, and majority-label aggregation erases the perspectives of the groups most affected \citep{haimson2021disproportionate, blodgett2020language}. These findings suggest that disagreement about severity reflects the subjective nature of the judgment and should not be treated as annotation failure.

\BfPara{Learning from annotator disagreement}
A growing body of work treats annotator disagreement as a signal \citep{aroyo2015truth, plank2022problem}. Disagreement persists in natural language inference even as more annotators are added \citep{pavlick2019inherent, nie2020learn}, some offensive-language corpora are released with their disagreement preserved \citep{leonardelli2021agreeing}, and taxonomies separate disagreement caused by ambiguity, sloppiness, and genuine subjectivity \citep{sandri2023why}. Surveys call for models that learn from the full label distribution, meaning the share of annotators who chose each label \citep{uma2021learning}. A soft label is one such distribution. It gives the share of annotators for each level (for example, 60\% moderate and 40\% severe) and not a single answer. Training on soft labels improves generalization and reduces overconfidence \citep{peterson2019human, fornaciari2021beyond}, and the LeWiDi shared task evaluates predicted distributions against annotator distributions \citep{leonardelli2023semeval}.

This line of work is known as data perspectivism \citep{basile2021perspectivist}. Within it, the prescriptive-descriptive distinction describes how labels are collected \citep{rottger2022two}. A prescriptive protocol gives annotators a rubric to apply, so the disagreement that remains is not caused by unclear instructions. A descriptive protocol asks annotators for their own subjective judgment. The dataset used in this study was annotated with a prescriptive rubric that gives examples, so the disagreement that survives is a consequence of the task's subjectivity and not of vague instructions.

Two approaches follow from this view. The first models individual annotators. Multi-annotator architectures predict each annotator's own labels \citep{davani2022dealing} and jury learning lets a designer choose whose voices a classifier follows \citep{gordon2021disagreement, gordon2022jury}. The advantage of these approaches is that they preserve each person's perspective and make the choice of whose judgment to follow explicit. Their cost is that they need annotator identities and several labels from each annotator, which most datasets do not record. The second approach models the overall distribution of labels for each comment. Its advantage is that it needs only the number of annotators who chose each label, which many datasets record. Its limit is that one model cannot say which annotator or group is right on a given comment. We study the second approach, because the dataset used here records label counts without annotator identities.

\BfPara{Moderation as human-AI collaboration}
Deployed moderation is a collaboration in which automated systems filter and humans decide the hard cases. Rule-based automation creates over- and under-flagging that humans must correct \citep{jhaver2019human}, moderators want to delegate conditionally while keeping uncertain cases for themselves \citep{lai2022human}, and current systems fail to reflect contextual assessments of harm \citep{gorwa2020algorithmic}. Formal treatments train the classifier and the rule for deferring a case to a human at the same time \citep{madras2018predict, mozannar2020consistent}. Previous work found that a model's own uncertainty can decide which cases reach a human, but it is not clear whether annotator disagreement would be a better signal for choosing which comments a moderator should see.

\BfPara{Calibration under disagreement}
Modern neural classifiers are often overconfident, meaning that their predicted confidence is higher than how often they are right. Temperature scaling is the standard post-hoc repair, and it divides the model's output scores by a single number fitted on held-out data \citep{guo2017calibration}. Label smoothing is an early form of soft-label training. It moves a small fixed amount of probability from the majority label equally onto all the other levels \citep{szegedy2016rethinking}. This share is the same for every comment, so it carries no information about how the annotators of a given comment actually divided. \citet{baan2022stop} argue that Expected Calibration Error (ECE), a widely used calibration metric, is unsound when humans disagree. ECE measures how far a model's confidence is from its accuracy against the majority label, and when annotators disagree, the majority label is not a valid reference point. They propose adjusting a model's probabilities toward the distribution of human judgments.

\BfPara{Open problems}
This literature leaves 3 problems open. First, it is unclear whether accuracy against the majority label can still separate models once annotators disagree or whether a different measure is needed. Second, it is unclear how much of the benefit of learning from annotator distributions requires retraining a model and how much a post-hoc adjustment of an existing model can provide. Third, it is unclear whether models that preserve the label distribution represent the judgments of different demographic groups equally well. We address these 3 problems in the subsequent sections of the paper.

%% file: sections/methods.tex
\section{Methods}
\label{sec:methods}

This section describes how we study what it takes to preserve annotator disagreement in a model that grades the severity of cyberbullying. We run 3 experiments, one for each research question introduced in the introduction (\Cref{fig:flow}). We keep each comparison controlled so that it changes 1 factor at a time. Experiment \ding{183} compares a majority-trained model with a distribution-trained model that differs from it only in its training target, and all 3 experiments share the same encoder and training setup, so any difference we find can be traced to the choice under study.
\begin{description}\setlength{\itemsep}{2pt}
\item[\ding{182}] \emph{Evaluation.} We score the same models in 2 ways, by accuracy against the majority label and by how closely a model's predicted distribution matches the annotators' distribution (\Pref{sec:protocol}). We run this experiment to test whether the 2 scores agree about which models are good.
\item[\ding{183}] \emph{Cost.} We compare a model trained on the annotators' label distributions with a model trained on the majority label and then adjusted after training (\Pref{sec:models} and \Pref{sec:spectrumnet}). We run this experiment to test how much of the benefit requires retraining.
\item[\ding{184}] \emph{Groups.} We repeat the comparison on a dataset that records annotator demographics (\Pref{sec:dataset}). We run this experiment to test whether preserving the distribution represents each group's judgments fairly.
\end{description}
All 3 experiments use the same training setup (\Pref{sec:setup}), so that differences between models come only from the choices we study.

\input{figures/fig_flow}

\BfPara[sec:dataset]{Dataset}
In this work, we use an extension of the University of Colorado Instagram dataset \citep{hosseinmardi2015detection}, which we call Insta-CTSR. It contains 106,618 comments from Instagram comment threads, annotated for cyberbullying and severity. Each comment was independently labeled by 5 Amazon Mechanical Turk\footnote{\url{https://www.mturk.com/}} Master Workers on an ordered 4-level scale of \emph{not cyberbullying}, \emph{mild}, \emph{moderate}, and \emph{severe}. A comment counts as cyberbullying when at least 3 of the 5 annotators place it above the not-cyberbullying level. This yields 11,310 cyberbullying comments, and the other 95,308 comments are not cyberbullying. The comments come from 430 Instagram comment threads that each contain at least one cyberbullying comment, with a median of 75 comments per thread. Because this corpus keeps all 5 labels per comment, it lets us study disagreement directly.

Of the 11,310 cyberbullying comments, 9,553 (84.5\%) have a unique plurality, meaning that one severity level receives strictly more labels than any other. These comments form our study set, and we use them for the majority-label comparisons, which need a single reference label. The remaining 1,757 (15.5\%) have no unique majority level, so we leave them out of the study set. We use them only in the held-out-label analysis (\Pref{app:heldout}), which needs no majority label.

We represent each comment's annotation as a label distribution over the 3 severity levels (mild, moderate, and severe). Because we score only comments that count as cyberbullying, we set the not-cyberbullying labels aside. We measure disagreement with its base-2 Shannon entropy \citep{shannon1948mathematical}, which is near zero under full agreement and rises as annotators diverge. \Cref{tab:dataset} shows how the 9,553 study-set comments are distributed across label patterns, which its caption explains. Only 392 comments (4.1\%) are fully unanimous, with all 5 annotators rating the comment as cyberbullying at the same severity level.

The disagreement holds under several ways of measuring it. We report Krippendorff's $\alpha$ \citep{hayes2007answering} with an ordinal distance, computed from the released per-annotator ballots, in 3 settings. Over all annotated comments and the full 4-level scale, $\alpha$ is 0.32. Annotators agree that most of these comments are not cyberbullying, which raises agreement. Over the 9,553 study-set comments and the 3 severity levels alone, $\alpha$ is 0.19. Over the same comments with the full 4-level scale, which keeps not cyberbullying as a fourth label, $\alpha$ is 0.09. The lower study-set values reflect the harder gradations that remain once the easy not-cyberbullying split is removed, and these comments stay contested no matter how the scale is defined.

The main experiments use this dataset. The generalization and group experiments of \Cref{sec:results} use 2 further datasets: EPIC, a binary irony dataset that records annotator gender, and ConvAbuse, a 5-level abuse-severity dataset. We use these additional datasets because our (Insta-CTSR) dataset records neither annotator demographics nor a second task and thus cannot show whether the findings hold elsewhere or how different groups are affected.

\input{tables/tab_dataset}

\BfPara[sec:setup]{Task and Setup}
The main task is severity grading. In this study, a model receives a cyberbullying comment together with the 5 comments that precede it in the same Instagram thread and predicts one of 3 severity levels: mild, moderate, or severe. The thread comments give the model the same context the human annotators saw when they judged the comment. The full 4-level scale, which adds not-cyberbullying as its own class, is used only in the encoder comparison of \Pref{app:backbone}, where a 4-class classifier with focal loss was used to stress the rare severe class when comparing encoders. All models share one setup so that any 2 models differ in exactly one design choice. Every model uses the DeBERTa-v3-base encoder \citep{he2021deberta}, which we selected by comparing 7 encoders under identical conditions (\Pref{app:backbone}), and models differ only in the prediction head placed on top of the encoder and in the training target. Every model trains with the same hyperparameters, listed in \Pref{app:training}, so a difference between 2 models reflects the design choice and not one model being tuned better than another. Evaluation uses 10-fold cross-validation \citep{stone1974cross} with folds built at the session level. An annotation session is the set of all comments on one Instagram post, so all comments from one session fall in the same fold and the model that scores a comment has never seen its session during training.

\BfPara[sec:models]{Model Families}
We evaluate whether common modeling choices for ordinal severity prediction lead to meaningful differences in performance. Our goal is not to identify a single best architecture, but to determine whether the patterns we observe are consistent across reasonable alternatives. To do so, we vary 3 aspects of the modeling setup: \emph{ordinal encoding}, \emph{task decomposition}, and \emph{conversational context and training data}. Our main reference model is \emph{Flat 3-class}. It is a standard softmax classifier that directly predicts the 3 severity levels (mild, moderate, and severe) using cross-entropy loss and majority labels. We compare the other models against this model.

We consider the following variants:

\ding{182} \emph{Ordinal encoding.} These models differ in how they represent the ordering among severity levels. CORAL and CORN replace the standard softmax output layer with cumulative-link formulations that explicitly model the ordering from mild to severe \citep{cao2020coral, shi2023corn}. EMD uses a different approach: it penalizes the squared distance between the predicted and target cumulative distributions, encouraging predictions that are close to the target in severity space \citep{hou2016squared}. Pairwise CORAL further adapts CORAL by learning severity through pairwise comparisons \citep{cao2020coral}.

\ding{183} \emph{Task decomposition.} We consider 2 approaches. The \emph{hierarchical} model splits the grading into 2 binary decisions: first, it separates mild from moderate-or-severe; second, among the comments predicted as moderate-or-severe, it separates moderate from severe. The \emph{detection-only} variant uses only the first decision and therefore distinguishes mild from more severe comments without predicting moderate versus severe.

\ding{184} \emph{Conversational context and training data.} We also test whether information outside the target comment affects severity prediction. The \emph{no-context-encoder} variant removes the context encoder, which allows us to measure the contribution of thread context without changing the underlying text encoder. The \emph{no-thread-context} variant retains the full hierarchical architecture but supplies an empty context input. This comparison helps distinguish the benefit of actual thread information from any benefit due simply to the larger model architecture. Finally, the \emph{CB-only} variant is trained only on cyberbullying comments and excludes the non-cyberbullying comments that are included in the main training set. These variants give us 11 models in the accuracy comparison (\Cref{tab:systems}), described in \Pref{sec:spectrumnet}. These models cover the main architectural and training choices that we expect practitioners and developers to consider when predicting ordinal severity.

For the fidelity analysis in \Pref{sec:calibration}, we additionally evaluate 3 models that are not included in the accuracy comparison. All 3 are defined in \Pref{sec:spectrumnet}: a flat model trained with KL divergence against the full label distribution and 2 soft-target baselines, label smoothing and SORD, which do not use the label distributions. These additional models allow us to separate improvements in predictive accuracy from improvements in fidelity to the underlying severity judgments. Because fidelity is the primary focus of our study, our main comparisons therefore center on the fidelity metrics.

\input{tables/tab_systems}
\BfPara[sec:spectrumnet]{Label-distribution training}
To examine whether a model trained to predict the label distribution outperforms a model that is calibrated after training, we need a distribution-trained model that differs from a standard one in only one way. We take the hierarchical model of \Pref{sec:models} and keep everything about it except the training target. The standard version learns from the majority label, which we call a hard label because it gives one answer per comment. We call this version the matched twin, since it differs from the distribution-trained model only in its target. The distribution-trained version learns from the annotators' split, a soft target that gives the share of annotators who chose each level, for example 60\% moderate and 40\% severe. Its first head learns the share of annotators who chose a level above mild and its second head learns the share of those who chose severe, both with binary cross-entropy. Since architecture, data, and hyperparameters are identical, any difference between the 2 versions comes from the training target alone.

Two questions could still be raised, so we built 2 additional models to answer them. The first question is whether the hierarchical design matters. A flat model trained with KL divergence against the full 3-way label distribution answers it, since it uses the label distribution without the hierarchy. The second question is whether any gain comes from the annotators' labels or simply from softer targets. Two models with softer targets answer it. Their targets are softened versions of the majority label, built without using how the annotators actually divided. Label smoothing moves a small share of probability ($\epsilon = 0.1$) off the majority label onto the other levels, and SORD places probability around the majority level according to its ordinal distance (\Pref{app:training}). In contrast, the distribution-trained model learns from the annotators' own labels for each comment. If the 2 softer-target models match the distribution-trained one, the gain should come from softening alone, and if they fall short, it should come from the annotators' labels.

\BfPara[sec:protocol]{Evaluation Protocol}
We score each model in 2 ways. The first asks how often the model picks the majority label, measured with macro-F1 so the rare severe class counts as much as the common ones. The second asks how closely the model's predicted split across mild, moderate, and severe matches the annotators' actual split, which we refer to as fidelity. We measure fidelity with total variation, the Brier score \citep{brier1950verification}, and the log score. The Brier score and the log score reward a model for honest probabilities and penalize it for confident mistakes \citep{gneiting2007strictly}. We use fidelity as our main measure, given our focus on disagreement, and macro-F1 is kept for comparison with prior work. The metrics used in this experiment are presented in \Pref{sec:metricdefs}.

The distribution-trained model has an unfair head start, because fidelity scores every model against the same label distributions it was trained on. We level the field in 2 ways. First, every model, including the baselines, gets the same calibration step: a 12-parameter map fitted to the label distributions of the training folds and applied to the held-out fold (\Pref{app:calibmaps}). After this step, each model has seen the label distributions once, so any remaining difference comes from training. Second, we hide one of the 5 labels per comment from both training and calibration and score every model against that hidden label (\Pref{app:heldout}). A model that only memorized the labels it saw will not predict the hidden one. As a reference point, we also fit a standard confidence calibration map to the majority labels alone \citep{guo2017calibration, kull2019beyond}. Statistical procedures are presented in \Pref{app:stats}.

We add 2 zero-shot large language models as external reference points that see none of the task-specific annotations and so are not competitive baselines. These are OpenAI's GPT-4o-mini through the OpenAI API and Qwen3.6-27B run locally for reproducibility. Each receives the comment, its thread context, and the annotation rubric, which gives one example comment for each severity level exactly as the human annotators saw it. Neither model receives demonstrations, i.e., additional labeled comments from our data that it could imitate. We read a severity distribution from the model's own probabilities that the next word is mild, moderate, or severe, rescaled to sum to 1 (\Pref{app:llm}).

%% file: figures/fig_flow.tex
\begin{figure}[!htbp]
\centering
\hyphenpenalty=10000\exhyphenpenalty=10000
\resizebox{0.8\textwidth}{!}{%
\begin{tikzpicture}[font=\small,
  pill/.style={fill=black,text=white,rounded corners=4pt,align=center,inner xsep=4pt,inner ysep=4pt,font=\footnotesize\bfseries,text width=2.3cm,minimum height=1.25cm},
  proc/.style={draw=black!70,fill=black!22,rounded corners=3pt,align=center,inner xsep=5pt,inner ysep=4pt,text width=5.0cm,minimum height=1.0cm,font=\footnotesize},
  proc2/.style={proc,text width=2.3cm,minimum height=1.25cm},
  ob/.style={draw=black!80,fill=white,rounded corners=3pt,minimum width=2.6cm,minimum height=2.2cm,anchor=north},
  dbox/.style={draw=black!70,fill=black!12,rounded corners=3pt,align=center,inner sep=4pt,text width=4.6cm,minimum height=0.7cm,font=\footnotesize},
  card/.style={draw=black!70,fill=white,rounded corners=3pt,align=left,inner xsep=6pt,inner ysep=4pt,text width=2.1cm,minimum height=1.25cm},
  panel/.style={draw=black!60,dashed,thick,rounded corners=5pt,fill=black!3},
  ttl/.style={font=\bfseries\small,anchor=north west},
  qst/.style={draw=black!40,fill=white,rounded corners=3pt,font=\itshape\footnotesize,text=black!85,align=center,inner xsep=5pt,inner ysep=4pt,text width=5.0cm,minimum height=1.45cm},
  lab/.style={font=\footnotesize,text width=2.3cm,align=center,anchor=north},
  arr/.style={-{Latex[length=2.2mm]},thick,black!75}]
\newcommand{\pp}[3]{\begin{scope}[shift={(#1,#2)}]\draw[line width=.5pt,fill=#3] (0,0.27) circle (0.12);\draw[line width=.5pt,fill=#3] (-0.17,-0.17) -- (-0.17,0.06) .. controls (-0.17,0.12) and (-0.1,0.13) .. (0,0.13) .. controls (0.1,0.13) and (0.17,0.12) .. (0.17,0.06) -- (0.17,-0.17) -- cycle;\end{scope}}
\newcommand{\pbub}[2]{\draw[line width=.6pt,fill=white,rounded corners=2pt] (#1,#2) rectangle ++(0.8,0.52);\draw[line width=.6pt,fill=white] (#1+0.18,#2) -- ++(-0.1,-0.17) -- ++(0.3,0.17);\draw[line width=.5pt] (#1+0.14,#2+0.18) -- ++(0.52,0);\draw[line width=.5pt] (#1+0.14,#2+0.34) -- ++(0.36,0);}
\newcommand{\pbars}[3]{\draw[fill=cmild,line width=.4pt] (#1,#2) rectangle ++(0.17,0.5);\draw[fill=cmod,line width=.4pt] (#1+0.24,#2) rectangle ++(0.17,0.25);\draw[fill=csev,line width=.4pt] (#1+0.48,#2) rectangle ++(0.17,0.38);}
\node[draw=black!60,thick,rounded corners=5pt,fill=black!3,minimum width=18cm,minimum height=3.9cm,anchor=north west] (top) at (0,0) {};
\node[ttl] at (0.3,-0.12) {STUDY DATA};
\node[draw=black!50,fill=white,rounded corners=4pt,minimum width=4.4cm,minimum height=1.9cm,anchor=north west] at (0.3,-0.75) {};
\pbub{0.55}{-1.75}
\pp{1.95}{-1.55}{cmild}\pp{2.45}{-1.55}{cmod}\pp{2.95}{-1.55}{cmod}\pp{3.45}{-1.55}{csev}\pp{3.95}{-1.55}{cmod}
\fill[cmild,draw=black,line width=.4pt] (1.7,-2.35) rectangle ++(0.14,0.14);\node[anchor=west,font=\scriptsize] at (1.84,-2.28) {mild};
\fill[cmod,draw=black,line width=.4pt] (2.5,-2.35) rectangle ++(0.14,0.14);\node[anchor=west,font=\scriptsize] at (2.64,-2.28) {moderate};
\fill[csev,draw=black,line width=.4pt] (3.6,-2.35) rectangle ++(0.14,0.14);\node[anchor=west,font=\scriptsize] at (3.74,-2.28) {severe};
\node[align=center,text width=4.4cm,anchor=north,font=\footnotesize] at (2.5,-2.75) {Comments, each labeled by multiple annotators};
\node[diamond,draw=black!70,fill=white,thick,aspect=1.9,align=center,inner sep=1pt,text width=1.9cm,font=\footnotesize] (d1) at (6.55,-1.6) {Counts as cyberbullying?};
\draw[arr] (4.7,-1.6) -- (d1.west);
\node[draw=black!50,fill=black!12,rounded corners=3pt,align=center,text width=2.0cm,font=\footnotesize,inner sep=3pt] (n1) at (6.55,-3.3) {Not scored};
\draw[arr,black!50] (d1.south) -- node[right,font=\footnotesize]{no} (n1.north);
\node[diamond,draw=black!70,fill=white,thick,aspect=1.9,align=center,inner sep=1pt,text width=1.9cm,font=\footnotesize] (d2) at (10.4,-1.6) {One level has the most labels?};
\draw[arr] (d1.east) -- node[above,font=\footnotesize]{yes} (d2.west);
\node[draw=black!50,fill=black!12,rounded corners=3pt,align=center,text width=2.2cm,font=\footnotesize,inner sep=3pt] (n2) at (10.4,-3.3) {Held-out-label analysis only};
\draw[arr,black!50] (d2.south) -- node[right,font=\footnotesize]{no} (n2.north);
\node[fill=black!22,draw=black!70,rounded corners=4pt,font=\small\bfseries,inner xsep=7pt,inner ysep=6pt] (study) at (13.2,-1.6) {Study set};
\draw[arr] (d2.east) -- node[above,font=\footnotesize]{yes} (study.west);
\node[card] (t1) at (16.3,-1.0) {};
\node[card] (t2) at (16.3,-2.8) {};
\fill[black] (15.25,-1.17) rectangle ++(0.3,0.3);
\node[anchor=west,align=left,text width=1.5cm,font=\footnotesize] at (15.7,-1.02) {Majority label};
\pbars{15.25}{-3.15}{}
\node[anchor=west,align=left,text width=1.5cm,font=\footnotesize] at (16.0,-2.9) {Label distribution};
\draw[arr] (study.east) -- (t1.west);
\draw[arr] (study.east) -- (t2.west);
\draw[thick,black!75] (8.4,-3.9) -- (8.4,-4.25);
\draw[thick,black!75] (2.9,-4.25) -- (9.0,-4.25);
\foreach \cx in {2.9,9.0}{\draw[arr] (\cx,-4.25) -- (\cx,-4.75);}
\foreach \x in {0,6.1,12.2}{\draw[panel] (\x,-4.75) rectangle ++(5.8,-9.95);}
\node[ttl] at (0.2,-4.9) {\ding{182}\ EVALUATION};
\node[qst] at (2.9,-6.15) {When annotators disagree, can accuracy against the majority label still tell a good model from a bad one?};
\node[dbox] (e1d) at (2.9,-7.5) {Study set};
\node[pill] (a1) at (1.5,-9) {Model families};
\node[pill] (a2) at (4.3,-9) {Zero-shot language models};
\draw[arr] (e1d.south -| a1) -- (a1.north);\draw[arr] (e1d.south -| a2) -- (a2.north);
\node[proc] (ap) at (2.9,-10.95) {each model scored in 2 ways};
\draw[arr] (a1) -- (a1 |- ap.north);\draw[arr] (a2) -- (a2 |- ap.north);
\node[ob] (o1) at (1.5,-12.2) {};
\node[ob] (o2) at (4.3,-12.2) {};
\draw[arr] (ap.south -| o1.north) -- (o1.north);\draw[arr] (ap.south -| o2.north) -- (o2.north);
\draw[line width=.6pt,fill=white] (1.5,-12.75) circle (0.25);\draw[line width=.6pt,fill=black] (1.5,-12.75) circle (0.1);
\node[lab] at (1.5,-13.1) {Accuracy against the majority label};
\pbars{3.97}{-13}{}
\node[lab] at (4.3,-13.1) {Fidelity to the annotators' label distribution};
\node[ttl] at (6.3,-4.9) {\ding{183}\ COST};
\node[qst] at (9.0,-6.15) {Can an existing model report the distribution of annotator labels without being retrained, and how much does retraining add?};
\node[dbox] (e2d) at (9.0,-7.5) {Study set};
\node[pill] (b1) at (7.6,-9) {Majority-trained model (matched twin)};
\node[pill] (b2) at (10.4,-9) {Distribution-trained model};
\draw[arr] (e2d.south -| b1) -- (b1.north);\draw[arr] (e2d.south -| b2) -- (b2.north);
\node[proc2] (b1p) at (7.6,-10.95) {Calibration step, no retraining};
\node[proc2] (b2p) at (10.4,-10.95) {Same calibration step};
\draw[arr] (b1) -- (b1p);\draw[arr] (b2) -- (b2p);
\node[ob] (p1) at (7.6,-12.2) {};
\node[ob] (p2) at (10.4,-12.2) {};
\draw[arr] (b1p) -- (p1.north);\draw[arr] (b2p) -- (p2.north);
\draw[line width=.6pt,fill=black!30] (7.25,-12.9) rectangle ++(0.19,0.18);\draw[line width=.6pt,fill=black!30] (7.5,-12.9) rectangle ++(0.19,0.36);\draw[line width=.6pt,fill=black!30] (7.75,-12.9) rectangle ++(0.19,0.54);
\node[lab] at (7.6,-13.1) {Gain from calibration alone};
\draw[line width=.6pt,fill=black!30] (10.05,-12.9) rectangle ++(0.19,0.5);\draw[line width=.6pt,fill=black!30] (10.3,-12.9) rectangle ++(0.19,0.5);\draw[line width=.6pt,fill=black] (10.3,-12.4) rectangle ++(0.19,0.14);
\node[lab] at (10.4,-13.1) {Further gain over the calibrated twin};
\node[ttl] at (12.4,-4.9) {\ding{184}\ GROUPS};
\node[qst] at (15.1,-6.15) {Does reporting that distribution represent each demographic group's judgments more fairly?};
\node[dbox] (e3d) at (15.1,-7.5) {EPIC irony dataset (records annotator gender)};
\node[pill] (c1) at (13.7,-9) {Majority-trained model};
\node[pill] (c2) at (16.5,-9) {Distribution-trained model};
\draw[arr] (e3d.south -| c1) -- (c1.north);\draw[arr] (e3d.south -| c2) -- (c2.north);
\node[proc] (cp) at (15.1,-10.95) {fidelity scored separately for each group};
\draw[arr] (c1) -- (c1 |- cp.north);\draw[arr] (c2) -- (c2 |- cp.north);
\node[ob] (q1) at (13.7,-12.2) {};
\node[ob] (q2) at (16.5,-12.2) {};
\draw[arr] (cp.south -| q1.north) -- (q1.north);\draw[arr] (cp.south -| q2.north) -- (q2.north);
\pp{13.4}{-12.75}{white}\pp{14.0}{-12.75}{black!55}
\node[lab] at (13.7,-13.1) {Gain for each group};
\draw[line width=.5pt,fill=white] (16.1,-12.9) rectangle ++(0.24,0.5);
\draw[line width=.5pt,fill=black!55] (16.5,-12.9) rectangle ++(0.24,0.2);
\draw[line width=.5pt,<->] (16.92,-12.7) -- (16.92,-12.4);
\node[lab] at (16.5,-13.1) {Gap between the groups};
\end{tikzpicture}}
\caption{Study design. Annotated comments are summarized as a majority label and as a label distribution (the colors of the annotator icons and bars stand for the severity levels mild, moderate, and severe). Three experiments follow, one for each research question of the introduction. Experiments 1 and 2 use the study set, and Experiment 3 uses EPIC, a second dataset that records annotator gender (our own dataset does not record annotator identities). Experiment 2 repeats its comparison on further datasets (not shown).}
\label{fig:flow}
\end{figure}
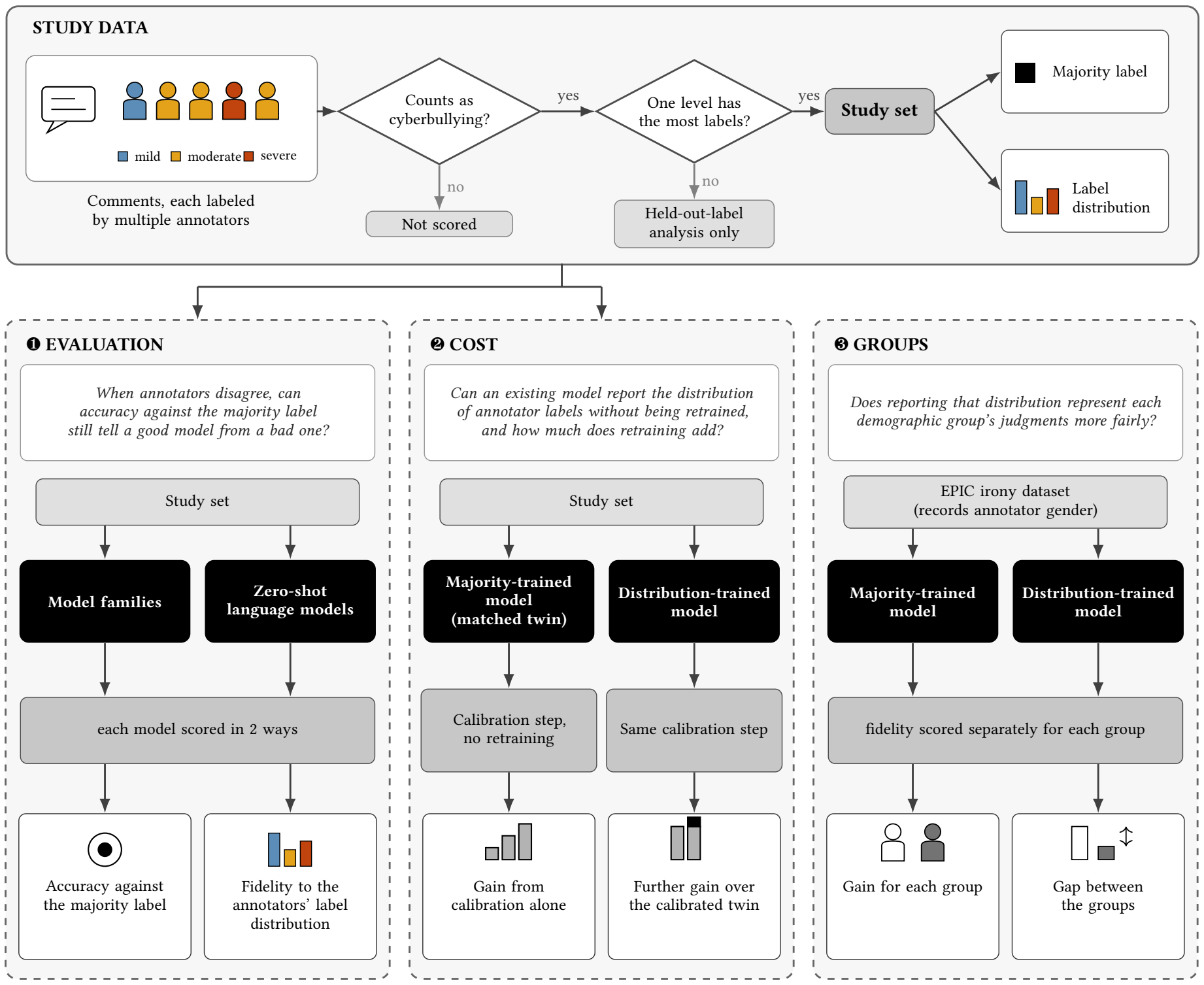

%% file: tables/tab_dataset.tex
\begin{table}[!htbp]
\centering
\begin{tabular}{lrrrr}
\toprule
Label pattern & Mild & Moderate & Severe & Total (\%) \\
\midrule
5--0 unanimous   & 121   & 192   & 79  & 392 (4.1) \\
4--1 strong      & 1,009 & 914   & 142 & 2,065 (21.6) \\
3--2 weak        & 1,407 & 1,237 & 90  & 2,734 (28.6) \\
3--1--1 plurality & 1,304 & 993   & 117 & 2,414 (25.3) \\
Split $\leq 2$   & 1,296 & 593   & 59  & 1,948 (20.4) \\
\midrule
All CB comments  & 5,137 (53.8\%) & 3,929 (41.1\%) & 487 (5.1\%) & 9,553 (100) \\
\bottomrule
\end{tabular}
\caption{The 9,553 study-set comments in Insta-CTSR by majority severity level and 5-annotator label pattern, with the percentage of all 9,553 comments in parentheses. A label pattern gives the number of annotators who chose each label, from the most to the least common. For example, 3--2 means that 3 annotators chose one label and 2 chose another, and 3--1--1 means that 3 annotators chose one label and the other 2 each chose a different label. Split $\leq 2$ means that no label was selected by more than 2 annotators.}
\label{tab:dataset}
\end{table}

%% file: tables/tab_systems.tex
\begin{table}[!htbp]
\centering
\footnotesize
\setlength{\tabcolsep}{4pt}
\begin{tabular}{p{0.27\textwidth}p{0.36\textwidth}p{0.27\textwidth}}
\toprule
System & What it does & Learns from \\
\midrule
Flat 3-class & Softmax head over mild, moderate, severe; the reference model & Majority label \\
CORAL head, CORN head & Output layer that respects the order mild < moderate < severe & Majority label \\
EMD loss & Penalizes the distance between predicted and target cumulative distributions & Majority label \\
Pairwise CORAL & Learns severity from pairwise comparisons & Majority label \\
Hierarchical & 2 binary decisions: mild vs.\ more severe, then moderate vs.\ severe & Majority label \\
Level-1 only (mild vs rest) & Only the first of those 2 decisions & Majority label \\
No context encoder & Flat classifier that reads the comment alone, with no context encoder & Majority label \\
No thread context & Hierarchical model given an empty context & Majority label \\
CB-only training & Trained on cyberbullying comments only, without non-cyberbullying comments & Majority label \\
Hierarchical (distribution-trained) & Hierarchical model whose targets are the annotators' split & Label distribution \\
\midrule
Flat KL (distribution-trained) & Flat model trained with KL divergence on the label distribution & Label distribution \\
Label smoothing & Moves a fixed amount of probability off the majority label & Majority label, softened \\
SORD & Places probability around the majority label by ordinal distance & Majority label, softened \\
\bottomrule
\end{tabular}
\caption{The systems we compare and what each learns from. The first 10 rows, with CORAL and CORN counted separately, form the 11 systems of the accuracy comparison. The last 3 rows are added for the fidelity comparison. Only the 2 distribution-trained systems see the annotators' label distributions.}
\label{tab:systems}
\end{table}

%% file: sections/results.tex
\section{Results}
\label{sec:results}

We take the 3 questions in order. The first asks whether accuracy can separate models (\Tref{sec:saturation}), the second how much of the benefit requires retraining (\Tref{sec:calibration} and \Tref{sec:generalization}), and the third whether the benefit reaches each annotator group (\Tref{sec:groups}). A final analysis asks whether model confidence can route comments to moderators (\Tref{sec:uncertainty}).

\BfPara[sec:saturation]{Accuracy Saturates Across Architectures}
\Cref{tab:ladder} reports macro-F1 for all model families under 10-fold cross-validation. The 5 strongest systems cluster between 0.68 and 0.70, and none of the ordinal formulations improves on that range: CORN reaches 0.674, EMD loss 0.665, CORAL 0.591, and Pairwise CORAL 0.587. The clustering is not an artifact of macro-F1, since the same systems also cluster under ordinal metrics, with quadratic-weighted $\kappa$, which measures agreement with the majority label and penalizes a prediction more the further it falls from that label on the mild-to-severe scale, near 0.50 to 0.55, and with mean absolute error, the average number of severity levels a prediction is off by, near 0.31. No point-estimate metric clearly separates the competitive systems from one another.

\input{tables/tab_ladder}

To see how good these numbers are, we treat each annotator as if they were a model. For each comment, we compare that annotator's severity label with the majority label of the other 4 annotators, compute macro-F1, and average across annotators. This gives a single-annotator reference of macro-F1 0.53. It measures how consistent one human is with the rest under the same procedure, so it is not a ceiling for a model, and the competitive systems sit well above it because they learn from aggregated labels. The reference is also not like-for-like, since each annotator is compared with the majority of 4 other labels while the models are compared with the majority of all 5 labels. The useful result is the narrow range among the competitive systems. Once a model is reasonably good, macro-F1 stops telling us which model most closely represents the human labels in this experiment.

The hierarchical distribution-trained model reaches 0.696 against 0.682 for the Flat 3-class model, with a bootstrap interval of $[+0.003, +0.028]$. The gain is real but small, and it is the only separation macro-F1 gives us across 11 systems. Thread context contributes about as much, between 0.005 and 0.033 in macro-F1 depending on how the context is removed (see \Pref{app:thread}).

This plateau is not a quirk of the encoder family. Two large language models prompted zero-shot with the annotators' rubric score below the competitive range (\Cref{tab:ladder}), with GPT-4o-mini at 0.443 macro-F1 and Qwen3.6-27B at 0.612. For a platform, this means that a higher majority-label accuracy does not show that a model represents the annotators' judgments better. Focusing on the majority label alone ignores important aspects of human judgments. We analyze these other aspects next.

\BfPara[sec:calibration]{Calibration Recovers Most of the Label Distribution}
\input{tables/tab_decomp}

\input{tables/tab_calibration}

We now ask how well models reproduce the annotators' label distribution. We measure this with 3 scores that compare a model's predicted distribution with the annotators' actual one: total variation (TV), the Brier score, and the log score (\Pref{sec:protocol} and \Pref{sec:metricdefs}). For each of them, lower is better and 0 is a perfect match. \Cref{tab:calibration} compares the 2 distribution-trained systems, the hierarchical model and the Flat KL model (a flat 3-class model trained to match the label distribution with KL divergence), with 3 baselines that never see the annotators' label distributions: the matched twin, label smoothing, and SORD. Label smoothing and SORD soften the majority label by a fixed rule and never use the annotators' own labels (\Pref{sec:spectrumnet}). Each system is shown before and after label calibration (\Pref{sec:protocol}).

\Cref{tab:decomp,tab:calibration} report one experiment at 2 levels of detail. \Cref{tab:calibration} compares all systems, and \Cref{tab:decomp} follows the matched twin step by step, starting from its raw score, which is its Brier score before any calibration, and ending at the best score any model could reach. To judge these numbers, we need to know how good a model could possibly be, and the answer is not perfect. The 5 labels for a comment are a sample of what annotators think, so even a model that knew exactly how the whole annotator pool would split could not match 5 sampled labels exactly. Simulation puts this sampling-noise floor at 0.069 Brier for this corpus (see details in \Pref{app:capacity}). The room for improvement therefore runs from the matched twin's raw Brier of 0.294 down to 0.069, and \Cref{tab:decomp} reports each step as a percentage of that range. The absolute Brier scores and the order of the steps are the evidence. The floor may also be generous for a model that reads only the text, which may not be able to reach it. A stricter floor would make every percentage larger while leaving the order unchanged.

The improvement in \Cref{tab:decomp} comes in 3 stages, each smaller than the one before. The first stage uses no label distributions at all. Temperature scaling divides the model's output scores by a single number fitted on the training data so that its confidence matches how often it is right. Fitted to the majority labels, it moves the matched twin from 0.294 to 0.182 Brier, closing about half the gap, and a richer map fitted to the majority labels reaches 0.164. All of this comes from fixing an overconfident model. The second stage adds the label distributions. A 12-parameter map fitted to them reaches 0.144 Brier with no retraining, an improvement of 0.018 over the majority-fitted map (bootstrap CI $[0.017, 0.020]$). The label distributions therefore carry information the majority label does not. The third stage trains on the label distributions directly and reaches 0.137 Brier, a further 0.007 (bootstrap CI $[0.005, 0.010]$). That is about a tenth of what remained, so training on the label distributions is a refinement on top of calibration.

The gain from training on the label distributions is small, but it is not noise. Retrained across 5 seeds and all 10 folds, the distribution-trained model beats its matched twin at every seed (\Pref{app:margins}). Its Brier advantage over the matched twin survives Holm correction across the family of fold-level tests, and the sign-consistency and held-out-label checks point the same way (\Pref{app:capacity} and \Pref{app:heldout}). The gain comes from the training target and not from the architecture. After label calibration, the flat KL control edges past \ours{} on every fidelity metric, and before calibration the 2 are level with \ours{} slightly ahead on Brier. What helps is learning the label distribution, in either architecture.

Fidelity depends on how the annotators split. Models do worst on unanimous comments, where all 5 annotators agree, with Brier scores of 0.32 to 0.34, and best on 3-to-2 comments, at 0.06 to 0.07. Although potentially counterintuitive, a soft-label model rarely predicts near-certainty, so it under-commits exactly where the annotators were certain. This highlights the benefit of calibration, since sharpening the predictions recovers the most on these unanimous comments. The distribution-trained model beats its matched twin on every label pattern.

The severe class behaves differently. On the 487 severe comments, the distribution-trained model has a slightly higher severe-class F1 before calibration (0.675 against 0.663) but, after calibration, the matched twin has the lower Brier in 6 of 10 folds (\Pref{app:capacity}). The gain from training on label distributions is therefore a mild-and-moderate-class effect.

Evaluation and deployment need different amounts of labeling. Measuring fidelity needs the full label distribution for every comment, but applying the map does not. Fitted from a single label per comment, the map reaches almost the same Brier as when fitted from all 5, 0.146 either way (\Cref{sec:discussion}), because its 12 parameters are estimated across the whole corpus and not per comment. A platform can therefore add the map to a severity model it already runs without retraining anything. The next paragraph checks whether these results depend on our corpus.

\BfPara[sec:generalization]{The Sources of Fidelity Gain Partly Generalize}
To test whether these results depend on our corpus, we repeat the decomposition on 2 additional corpora that keep per-annotator labels. EPIC \citep{frenda2023epic} is a binary irony task and ConvAbuse \citep{cercas2021convabuse} is a 5-level abuse-severity task. On each, we fine-tune the same DeBERTa-v3-base encoder on majority labels, apply the same sequence of steps, and read the result against that corpus's own sampling-noise floor (\Cref{tab:generalize}).

Part of the decomposition holds on every corpus and part of it does not. We use the 3 steps defined in \Pref{sec:calibration}: temperature scaling, which uses no label distributions, the label-distribution map, which is fitted to them but applied to an already trained model, and training on the label distributions, which retrains the model. Temperature scaling recovers most of the closable gap on all 3 corpora using no label distributions, reaching 63\% on EPIC and 27\% on ConvAbuse against 50\% on our corpus. Fixing overconfidence is the dominant step everywhere, and how far it gets tracks how overconfident the base model was.

The extra gain from the label-distribution map appears only on our corpus, where it lifts the recovered fraction from 50\% to 67\%. On the replication corpora, it adds nothing over temperature scaling (63\% against 63\% on EPIC, and 22\% against 27\% on ConvAbuse). The multi-parameter map step is therefore specific to the severity task, while temperature scaling generalizes across all three. Training on the label distributions behaves the opposite way and adds a small consistent step everywhere, a further 3\% of the gap on our corpus, 4\% on EPIC, and 5\% on ConvAbuse. The replications support the paper's central claim, that most of the recoverable fidelity comes from calibrating a finished model, while showing that which form of calibration suffices depends on the corpus.

\input{tables/tab_generalize}

\BfPara[sec:groups]{Preserving the Distribution Helps Both Groups but Does Not Close the Gap}
Our corpus records label counts but not annotator identities, so we answer the third question on EPIC, which records annotator gender (\Pref{sec:dataset}). On EPIC, men and women assign nearly the same overall irony rate (0.323 against 0.304) yet flip the majority call on 26\% of items, so pooling to a single label erases a group-level split rather than a shared judgment. We score each trained model's predictions against the label distribution of men and of women separately, over 10 folds and 5 training seeds (\Cref{tab:equity}).

Three findings stand out. First, a majority-trained model represents both groups poorly and represents women worse than men, at Brier 0.318 against 0.283. Second, preserving the label distribution improves both groups over the calibrated majority baseline, and the improvement is significant for each: men by 0.007 in Brier (95\% CI $[0.003, 0.010]$, Wilcoxon $p = 0.0002$) and women by 0.005 (CI $[0.002, 0.008]$, $p = 0.003$) across the 50 fold-by-seed runs. Third, the improvement does not narrow the gap between the groups. The women-minus-men gap is 0.037 under the calibrated majority model and 0.038 under the preserving one, a difference of $+0.0015$ with a confidence interval that spans zero (CI $[-0.002, +0.005]$, $p = 0.44$).

\input{tables/tab_equity}

\BfPara[sec:uncertainty]{Additional Analysis: Model Confidence Does Not Track Human Disagreement}
It would be convenient if models were uncertain exactly where humans disagree, but they are not. We compare model confidence, the highest predicted class probability, against annotator disagreement, measured as label entropy. The relationship is weak: Spearman's $\rho$ is 0.117 for \ours{} and 0.116 for the Flat 3-class model, and both are statistically significant only because the sample is large. The LLM reference points behave the same way, at $-0.10$ for GPT-4o-mini and 0.15 for Qwen3.6-27B.

A more pointed question is whether model uncertainty can pick out contested comments at all. Used as a detector, model entropy reaches an area under the ROC curve of 0.54, where 0.50 is a coin flip. Calibration does not fix this, since after label calibration, the correlation moves only from 0.117 to 0.143 and the detector stays at chance. Calibration improves fidelity to the label distribution across the corpus, while it does little for identifying which individual comments will divide annotators.

Errors do concentrate where people disagree. We call a comment contested when no severity level receives at least 4 of the 5 labels. These are about 3 quarters of the study set and their error rate is 1.9 times that of clear comments (\Pref{app:errorconc}). A cyberbullying comment carries at most 2 not-cyberbullying labels, so any level reaching 4 labels must be mild, moderate, or severe, which makes this definition match the contested rows of \Cref{tab:dataset} exactly. A model can therefore fail on contested comments without being able to flag them ahead of time.

\BfPara[sec:deferral]{Deferral}
The obvious use of uncertainty is deferral, where the comments the model is least confident about go to human moderators and the model handles the rest \citep{madras2018predict, mozannar2020consistent}. We rank comments 3 ways: by model confidence, by an oracle that knows the true human label entropy, and by combining both. We also vary the proportion of comments sent to humans, from 10\% to 50\%. We report accuracy on the comments the model keeps, since the question is how often the automated decisions are right after routing, and per-class F1 grows unstable as the kept set shrinks.

Within realistic budget constraints, knowing where humans disagree does not significantly improve performance. At 30\% deferral, the combined ranking reaches 0.772 accuracy against 0.770 for confidence alone, and the difference is low (the bootstrap interval includes zero). Even an oracle with the true label entropy does not beat the model's own confidence.

At 50\% deferral, the picture looks different, with the combined ranking at 0.828 against 0.797, but the gain is confounded. Deferring the highest-entropy comments removes exactly those whose majority labels are least reliable, so part of the improvement comes from an easier evaluation set and not from better routing (\Pref{app:deferral}). We treat 30\% as the operative setting, since human review is the scarce resource in moderation.

For fidelity, we do not use label entropy as a routing signal at all. High-entropy comments are exactly where soft-label models under-commit and score worst (\Pref{sec:calibration}), so routing them away would change the kept set in a way that reflects the model's calibration and not the quality of the routing. The practical split is to route by model confidence and to use the label distribution to describe how divided the annotators are.

%% file: tables/tab_ladder.tex
\begin{table}[!htbp]
\centering
\begin{tabular}{lc}
\toprule
Model & Macro-F1 \\
\midrule
Level-1 only (mild vs rest)     & 0.452 {\scriptsize $\pm\,0.012$} \\
Pairwise CORAL                  & 0.587 {\scriptsize $\pm\,0.020$} \\
CORAL head                      & 0.591 {\scriptsize $\pm\,0.032$} \\
No context encoder              & 0.652 {\scriptsize $\pm\,0.016$} \\
EMD loss                        & 0.665 {\scriptsize $\pm\,0.031$} \\
CORN head                       & 0.674 {\scriptsize $\pm\,0.026$} \\
No thread context               & 0.680 {\scriptsize $\pm\,0.026$} \\
CB-only training                & 0.680 {\scriptsize $\pm\,0.024$} \\
Flat 3-class                    & 0.682 {\scriptsize $\pm\,0.020$} \\
Hierarchical                    & 0.685 {\scriptsize $\pm\,0.019$} \\
Hierarchical (distribution-trained)    & 0.696 {\scriptsize $\pm\,0.021$} \\
\midrule
GPT-4o-mini (zero-shot)         & 0.443 \\
Qwen3.6-27B (zero-shot)         & 0.612 \\
\bottomrule
\end{tabular}
\caption{Macro-F1 (mean $\pm$ SD over 10 folds) for each system. Every system except the distribution-trained one learns from the majority label. The 5 strongest systems fall between 0.68 and 0.70, so macro-F1 barely separates them. For reference, a single annotator compared with the majority label of the other 4 annotators reaches 0.53, and the 2 zero-shot language models (single runs) are shown at the bottom.}
\label{tab:ladder}
\end{table}

%% file: tables/tab_decomp.tex
\begin{table*}[t]
\centering
\setlength{\tabcolsep}{4pt}
\begin{tabular}{lcccr}
\toprule
Step (applied to the matched twin) & TV & Brier & Log & \% gap \\
\midrule
Raw (trained on majority labels)   & 0.318 & 0.294 & 1.244 & 0 \\
~~+ temperature, no label distributions & 0.263 & 0.182 & 0.836 & 50 \\
~~+ majority map, no label distributions & 0.237 & 0.164 & 0.798 & 58 \\
~~+ label-distribution map (12 parameters) & 0.233 & 0.144 & 0.764 & 67 \\
~~+ distribution training           & 0.228 & 0.137 & 0.755 & 70 \\
\midrule
Sampling-noise floor               & --    & 0.069 & --    & 100 \\
\bottomrule
\end{tabular}
\caption{Where the fit to the label distributions comes from. Each step adds the information named to the step above it, starting from the matched twin, the hierarchical model trained on the majority label, and moving toward the sampling-noise floor. TV is total variation, Brier is the Brier score, and Log is the log score. Each measures how far the predicted distribution is from the human label distribution, so lower is better. The last column is the percentage of the closable Brier gap ($0.294 \rightarrow 0.069$) that each step reaches. Temperature scaling closes half the gap using no label distributions. A small map fitted to the label distributions closes most of the rest without retraining. Training on the label distributions adds a little more.}
\label{tab:decomp}
\end{table*}

%% file: tables/tab_calibration.tex
\begin{table*}[t]
\centering
\setlength{\tabcolsep}{5pt}
\newcommand{\sd}[1]{{\scriptsize$\pm#1$}}
\begin{tabular}{lcccc}
\toprule
Model & TV & Brier & Log score & F1 \\
\midrule
Hierarchical (majority label, matched twin) & 0.318\,\sd{.009} & 0.294\,\sd{.019} & 1.244\,\sd{.045} & 0.685\,\sd{.019} \\
~~+ label calibration      & 0.233\,\sd{.016} & 0.144\,\sd{.019} & 0.764\,\sd{.024} & 0.682\,\sd{.026} \\
Label smoothing, $\epsilon = 0.1$ (softened majority label) & 0.319\,\sd{.008} & 0.271\,\sd{.016} & 1.116\,\sd{.023} & 0.690\,\sd{.025} \\
~~+ label calibration      & 0.231\,\sd{.017} & 0.143\,\sd{.018} & 0.761\,\sd{.025} & 0.683\,\sd{.025} \\
SORD (softened majority label) & 0.280\,\sd{.016} & 0.167\,\sd{.016} & 0.847\,\sd{.012} & 0.695\,\sd{.027} \\
~~+ label calibration      & 0.231\,\sd{.018} & 0.143\,\sd{.019} & 0.760\,\sd{.025} & 0.693\,\sd{.024} \\
Hierarchical (distribution-trained) & 0.233\,\sd{.014} & 0.155\,\sd{.015} & 0.794\,\sd{.027} & \textbf{0.696}\,\sd{.020} \\
~~+ label calibration      & 0.228\,\sd{.018} & 0.137\,\sd{.021} & 0.755\,\sd{.028} & 0.693\,\sd{.024} \\
Flat KL (distribution-trained)     & 0.233\,\sd{.016} & 0.157\,\sd{.018} & 0.793\,\sd{.031} & \textbf{0.696}\,\sd{.025} \\
~~+ label calibration      & \textbf{0.226}\,\sd{.018} & \textbf{0.136}\,\sd{.019} & \textbf{0.751}\,\sd{.025} & 0.695\,\sd{.023} \\
\bottomrule
\end{tabular}
\caption{Fidelity to the human label distribution (10-fold means $\pm$ SD, lower is better except F1, with the best mean in bold). F1 is macro-F1 of the most-probable label, reported for the calibrated rows as well because calibration can change which label is most probable, though here it rarely does. Label calibration is a Dirichlet map fitted on training-fold label distributions and applied cross-fold.}
\label{tab:calibration}
\end{table*}

%% file: tables/tab_generalize.tex
\begin{table}[!htbp]
\centering
\setlength{\tabcolsep}{4pt}
\begin{tabular}{llccc}
\toprule
Corpus & Task & + temperature & + label-distribution map & + distribution training \\
\midrule
Insta-CTSR (ours) & Severity, 3-class & 50 & 67 & 70 \\
EPIC              & Irony, binary     & 63 & 63 & 67 \\
ConvAbuse         & Abuse, 5-class    & 27 & 22 & 32 \\
\bottomrule
\end{tabular}
\caption{The decomposition across 3 disagreement corpora, as the percentage of the closable Brier gap reached after temperature scaling with no label distributions, a map fitted to the label distributions, and training on the label distributions. The replication values are means over 5 seeds, with a per-seed SD of at most 4 points. Temperature scaling recovers most of the gap everywhere. The further gain from the label-distribution map appears only on Insta-CTSR. Training adds a small consistent step on all 3.}
\label{tab:generalize}
\end{table}

%% file: tables/tab_equity.tex
\begin{table}[!htbp]
\centering
\footnotesize
\setlength{\tabcolsep}{6pt}
\begin{tabular}{lcc}
\toprule
Model (EPIC) & Men's labels & Women's labels \\
\midrule
Majority-trained          & 0.283 & 0.318 \\
~~+ label calibration     & 0.212 & 0.249 \\
Label-preserving          & 0.205 & 0.244 \\
\bottomrule
\end{tabular}
\caption{How well each model reproduces each group's judgments on EPIC, scored as Brier against the label distribution of men and of women separately (lower is better, means over 10 folds and 5 seeds). Preserving the label distribution improves both groups significantly over the calibrated majority baseline, but the gap between them does not narrow (0.037 against 0.038, difference not significant).}
\label{tab:equity}
\end{table}

%% file: sections/discussion.tex
\section{Discussion}
\label{sec:discussion}

\BfPara{A design recommendation and its evidence boundary}
We frame what follows as a design recommendation rather than a validated deployment, since we test model behavior and not moderator outcomes. Our results separate 2 jobs that pipelines usually run together. For the first job, deciding which comments a human reviews, the model's own confidence is already the right signal (\Pref{sec:deferral}). For the second, telling the human what is known about a comment, a single label is a poor summary. A comment that 3 annotators rate moderate and 2 rate severe and a comment that all 5 rate moderate call for different handling, and the distribution-trained models represent that difference best (\Pref{sec:calibration}). The design we recommend is a split: route by confidence and make the predicted distribution available. In practice, this needs no new infrastructure. A platform keeps the severity model it already runs, fits a 12-parameter map to the labels it already collects, reports the resulting distribution to moderators, and ranks what reaches them by model confidence. Our evidence stops at what the model produces. Whether a displayed distribution actually improves moderator decisions is an open question for human-factors research, since a distribution can bias a moderator as easily as it can inform them \citep{lai2022human}.

\BfPara{What multi-annotator budgets buy}
Collecting 5 labels per comment costs 5 times as much as collecting one \citep{sheng2008get}. The expense buys neither accuracy, which the majority-label metric cannot resolve among competitive systems, nor better triage. It buys the shape of human judgment and the consequences are concrete. The calibrated distribution-trained model and its calibrated twin assign a different most-probable label on 19\% of comments, but disagree on whether a comment is severe, the decision that triggers escalation, on only 2\%. The twin flags more comments as severe than the distribution-trained model does (4.4\% against 3.5\%; \Pref{app:stats}). The label differences sit in the mild-and-moderate range, consistent with the severe class being where training on the label distribution does not help.

Most of this value is available at low cost. It comes through a calibration map fitted to the label distributions and added to any existing model (\Pref{sec:calibration}), with a small further margin from training on the label distributions directly. The case here is one of adoption cost rather than a demonstrated moderator gain. Fitting that map on labels a platform already collects is close to free, adds no model retraining and no serving cost, and can sit on top of a severity model already in production. The barrier to reporting a distribution instead of a single label is therefore low, even though its downstream benefit remains to be tested.

The map barely needs multiple labels at all. Fitted to a single label per comment, it reaches the same fidelity as fitting it to all 5 (Brier 0.146 either way), because its 12 parameters are estimated across the whole corpus rather than per comment. A platform can then calibrate to disagreement using the single labels it already has, and teams choosing between annotating more items once and fewer items several times can weigh that trade directly. This calibrate-first lesson is not specific to cyberbullying. On any task whose labels are subjectively contested, calibrating an existing model recovers most of the distributional signal, and training on the label distributions adds a small refinement. Which form of calibration suffices varies by corpus: on our Insta-CTSR severity data, a label-distribution map is needed to reach the full gain, while on the replication corpora, plain temperature scaling already recovers most of it (\Pref{sec:generalization}).

\BfPara{Equity}
The group results (\Pref{sec:groups}) are what a shared model would be expected to produce. Because a single model shares one representation across annotators, it cannot encode which group is right on a given item, a limit we described in \Cref{sec:background}, so distribution preservation lifts fidelity to each group without favoring either. It is necessary for group representation but not sufficient to equalize it, and closing the gap needs the per-annotator and jury methods \citep{davani2022dealing, gordon2022jury, fleisig2023majority} that this line of work builds toward. On our own corpus, gender-targeted attacks generate more disagreement between annotators than any other topic (\Pref{app:disagreement}), and training on the label distribution keeps the outvoted severe labels in the predicted distribution, whereas majority training erases them. Whether specific communities' judgments are preserved depends on having annotator identities, as used in multi-annotator modeling \citep{davani2022dealing} and jury learning \citep{gordon2022jury}, which our corpus lacks \citep{haimson2021disproportionate, waseem2016racist}.
\input{tables/tab_shared}
\Cref{tab:shared} presents the advantages and limits of one shared model.

We still consider this a limited result, as it is based on a single binary group split in a lower-stakes irony task instead of a task involving harm. Part of the raw-Brier gap can also be a floor effect rather than under-representation, since a group whose labels are sparser or more dispersed has a higher sampling floor, and fidelity to a group's labels measures how well the model represents that group's judgments, not how much harm the group experiences. A fuller treatment needs more groups, gaps normalized against each group's own floor, and a corpus where the task carries real stakes. Although preserving the distribution does not settle whose judgment should prevail or equalize whose judgment is captured, it does prevent the training pipeline from silently erasing either.

\BfPara{Limitations and future work}
Our main evidence comes from one corpus, one platform, and one language, though the calibrate-first result replicates on 2 additional disagreement corpora that span a different task and label scheme (\Pref{sec:generalization}). The replications use the same fine-tuning configuration as our main runs rather than one tuned per corpus, are averaged over 5 seeds, and all 3 corpora are in English. Furthermore, the comments date from Instagram between 2013 and 2015 and were annotated by crowd workers rather than professional moderators or affected users, so the topic-level findings are era-specific. These third-party annotators may also judge severity differently from the targets whose experience the labels ultimately reflect. Instead of filtering labels based on worker reliability, as methods like MACE \citep{hovy2013learning} would enable, we treat the label distribution as something to be reproduced, given that filtering would, itself, impose a particular view of whose judgments are considered valid. Limitations of the present work thus underscore the need for future efforts that extend the comparison to other platforms, languages, contexts in which judgments of harm have significant downstream consequences, and more recent social media data.

Several limitations concern the method and its results. The training-signal margin is not uniform across the scale. On the 487 severe comments that the escalation use case depends on most, the benefit does not carry, and the twin is slightly ahead by only a small margin, near 0.006 Brier in 6 of 10 folds (\Pref{app:capacity}). Thus, the benefit we report is a mild-and-moderate-class effect. The association between human disagreement and model error is partly mechanical, because the majority label is fragile where labels split. The LLM reference points are 2 models queried zero-shot. They bound off-the-shelf grading rather than what tuned frontier models might reach. In light of this, one promising direction for future work involves prompting or tuning frontier models with the annotators' label distributions to test how far off-the-shelf grading can be raised.

Finally, a study with human moderators could test whether a displayed prediction distribution changes decisions on contested comments, which would connect this work to conditional delegation \citep{lai2022human}. Such a study should include moderators with varying identities and demographics, because their own judgments of severity may differ in the same ways that the annotators' do.

\BfPara{Code and data}
The code, the annotated corpus with all 5 labels per comment, the fold split and the per-system predictions are released at \coderepo.

%% file: tables/tab_shared.tex
\begin{table}[!htbp]
\centering
\footnotesize
\setlength{\tabcolsep}{4pt}
\begin{tabular}{p{0.20\textwidth}p{0.36\textwidth}p{0.36\textwidth}}
\toprule
Approach & Pros & Cons \\
\midrule
One shared model (tested here)
& Improves the fit to both groups (0.007 Brier for men, 0.005 for women). Needs only label counts, with no annotator identities. Adds only a calibration step. Favors neither group.
& Cannot encode which group is right on a given comment. Leaves the gap between the groups unchanged (0.037 against 0.038). Represents the pooled split, so differences between groups stay hidden. \\
\addlinespace
Per-annotator or jury models (not tested here)
& Can represent each annotator's or group's judgment and let the designer choose whose judgment to follow \citep{davani2022dealing, gordon2022jury}.
& Need annotator identities, which this corpus does not record. Leave the choice of whose judgment prevails to the designer. \\
\bottomrule
\end{tabular}
\caption{One shared model compared with per-annotator and jury models. The first row summarizes our EPIC results. The second row describes approaches from prior work that we did not test.}
\label{tab:shared}
\end{table}

%% file: sections/appendix.tex
\appendix

\section{Technical Appendix}
\label{sec:appendix}

\BfPara[app:ethics]{Ethical statement}
The corpus is public Instagram content used in a manner consistent with the source dataset's terms \citep{hosseinmardi2015detection}, with usernames pseudonymized before annotation. Severity annotations were collected under institutional review board approval from MTurk Master Workers screened for a 95\% approval history and US residency. Workers were paid \$0.042 per comment, a rate consistent with platform norms. Attention checks were applied, and failures were blocked and re-annotated \citep{instactsr}. LLM queries used a commercial API under its no-training terms, on pseudonymized text only. The few comments we quote are masked, stripped of user mentions, and already in the dataset's governed public release, and we release no personally identifying information. Systems built on this work grade the severity of speech, and errors carry asymmetric costs, since over-grading suppresses legitimate expression while under-grading leaves victims unprotected. Reporting the predicted distribution instead of a single label keeps the contested status of a comment visible to the human who acts on it. We do not recommend fully automated enforcement based on these models.

\BfPara[app:backbone]{Backbone selection}
The DeBERTa-v3-base backbone was fixed by comparing 7 encoders under one shared training configuration on the same session-level folds. This selection stage used a 4-class flat classifier with focal loss ($\gamma = 2.0$) to stress the rare severe class during the comparison. The ladder itself uses the 3-class severity task and the training losses stated in \Pref{sec:models}, with no focal loss. \Cref{tab:backbones} reports the full comparison. DeBERTa-v3-base gave the best macro-F1, the lowest mean absolute error and the lowest cross-fold variance, and it tied for the best quadratic-weighted $\kappa$ (QWK). It also beat the hate- and toxicity-pretrained encoders (HateBERT, ToxicBERT, Twitter-RoBERTa), so we fixed it for the whole ladder. The 5-comment context window changed QWK by only $+0.005$ at higher variance. The 0.617 to 0.652 macro-F1 range across encoders is itself early evidence of the accuracy ceiling this paper documents.

\BfPara[app:training]{Training}
All ladder systems fine-tune DeBERTa-v3-base with the comment and its 5 preceding thread comments encoded separately and combined before the head. Comments are truncated at 64 tokens. Every run uses 8 epochs, batch size 32, AdamW at a learning rate of $2\times10^{-5}$ and gradient clipping at 1.0. Folds are fixed once at the session level and shared by every system.

Seeds are used in 3 places. The main ladder of \Pref{sec:saturation} and \Pref{sec:calibration} uses seed 42 for data order and initialization, so that every system in the ladder sees the same conditions. The seed check of \Pref{app:margins} retrains \ours{} and its matched twin across all 10 folds under training seeds 1 through 5, with the data split unchanged. The replications of \Pref{sec:generalization} also run 5 seeds per corpus under this same configuration.

The label-smoothing control uses $\epsilon = 0.1$. The SORD control builds its target for a comment with majority class $y$ as $\mathrm{softmax}(-|k-y|/\tau)$ over classes $k$ with $\tau = 1$. All fine-tuning ran under Linux on NVIDIA A100 (40\,GB) GPUs, using PyTorch 2.7 and Transformers 4.45.2. The classical and frozen-feature baselines ran on NVIDIA Quadro RTX 8000 (48\,GB) GPUs under the same software stack.

\BfPara[app:calibmaps]{Calibration maps}
The label-fitted map is full Dirichlet calibration, $q = \mathrm{softmax}(W \log p + b)$ with $W \in \mathbb{R}^{3\times3}$ and $b \in \mathbb{R}^{3}$, so 12 parameters in total. It is initialized at the identity and fitted with Nelder-Mead (at most 8,000 iterations) to minimize the mean cross-entropy between $q$ and the severity-label distribution, with probabilities clipped at $10^{-12}$ inside the fit. The predictions entering each fit are out-of-fold predictions on the training folds, and the fitted map is applied to the held-out fold.

The majority-fitted temperature, vector and Dirichlet maps minimize negative log-likelihood against majority labels instead. These majority-fitted maps are fitted per fold on that fold's own test data, which is the most favorable data a post-hoc method can receive, and we use them only as a confidence-calibration baseline. All log scores in the paper clip predicted probabilities at $10^{-9}$.

\BfPara[sec:metricdefs]{Metric definitions in full}
The F1 score for one class is the harmonic mean of precision and recall on that class, so it rewards finding a class without over-firing on it. Macro-F1 is the unweighted mean of the per-class F1 scores against the majority label, which keeps the rare severe class from being drowned out by the common ones.

For the fidelity metrics, take a comment with predicted severity distribution $p$ and human label distribution $q$ over the 3 severity levels. Total variation is $\mathrm{TV}(p,q) = \tfrac{1}{2}\sum_k |p_k - q_k|$, the smallest amount of probability mass one must move to turn one distribution into the other. The Brier score is $\sum_k (p_k - q_k)^2$, the sum of squared differences between the two. The log score is $-\sum_k q_k \log p_k$, the negative log probability the model assigned to the outcomes. Log scores clip predicted probabilities at $10^{-9}$ before scoring, which matches the clipping used in the calibration fits. All fidelity metrics normalize $p$ and $q$ over the 3 severity counts alone, so not-cyberbullying labels are dropped and never reassigned to mild.

\input{tables/tab_backbones}

\BfPara[app:heldout]{Held-out-label protocol}
For each majority-CB comment, one severity label is removed by a fixed-seed generator (NumPy \texttt{default\_rng(42)}) over a fixed row order. Only severity labels are ever held out, and the 1,757 no-plurality comments never enter any training set. \Ours{} retrains on distributions over the remaining labels and the twin retrains on their majority with ties broken toward the more severe level. The calibration map refits on remaining-label distributions, and both pipelines are then scored against the single held-out label, which neither of them has seen.

\BfPara[app:llm]{LLM queries}
Both LLMs receive the system prompt below, which reproduces the human annotators' instructions and examples verbatim, followed by a user message containing the preceding thread comments (truncated at 2,000 characters) and the target comment (1,000 characters). GPT-4o-mini was queried through the chat completions API in July 2026 with \texttt{max\_tokens} 1, temperature 0 and \texttt{top\_logprobs} 15, and it answers with one word. Qwen3.6-27B is the 27B variant of Qwen3.6, run locally from the public ollama library (\texttt{ollama pull qwen3.6:27b}, digest \texttt{a50eda8ed977}, model card \url{https://ollama.com/library/qwen3.6}) with thinking disabled and the same log-probability settings. It answers with one digit (1 mild, 2 moderate, 3 severe) because its tokenizer splits the class words ambiguously.

Each severity distribution is the softmax over the maximum log probability observed for each class token, and a class absent from the returned top-15 list receives zero mass. This zero-mass clipping fires on at least one class for 97\% of comments for GPT-4o-mini, which rarely places every severity word in its top-15 list, and for none of the comments for Qwen3.6-27B, which returns full distributions. The GPT-4o-mini log score is therefore dominated by the clip constant and not by calibration, so we read its log-score comparison with caution and rely on total variation and the Brier score for the language-model references. Only pseudonymized comment text left the machine, and only for the API model.

\begin{quote}\small
You are annotating Instagram comments. Given the preceding thread comments as context and one target comment, select the severity of the cyberbullying present in the target comment.

Mild: Using mildly offensive language to attack or embarrass another person. Example: ``You're an idiot, why would you do that?''

Moderate: Intense, offensive language which includes attacking personal beliefs or core values. Example: ``I bet you voted Trump you stupid asshole''

Severe: Communicating physical threats, self-harm, blackmail, sexual assault, stalking, etc. Example: ``You should kill yourself before I find you and slash your throat in front of your kids''

Answer with exactly one word: mild, moderate, or severe.
\end{quote}

\BfPara[app:stats]{Statistical procedures}
Bootstrap intervals over fold differences use 10,000 resamples. Comment-level bootstraps in the deferral analysis are complemented by session-level cluster bootstraps with 2,000 resamples. Rank correlations are Spearman, group comparisons are Mann-Whitney, and multiple-comparison control is Holm over the stated family of tests.

The label-change rates quoted in \Cref{sec:discussion} compare the calibrated \ours{} with its calibrated twin on the same out-of-fold predictions. The 2 assign a different most-probable label on 19\% of study-set comments and disagree on the severe-versus-not escalation flag on 2\%. The twin flags severe on 4.4\% of comments against 3.5\% for \ours{}, so the twin is the one that over-flags severe, and 1.4 of the 2.0 points of flag disagreement are comments the twin calls severe and \ours{} does not.

\BfPara[app:thread]{Thread contribution}
Two ladder entries measure the thread differently and should be read together. The no-context-encoder variant (0.652) removes the context encoder entirely, while the no-context ablation (0.680) keeps the dual-encoder architecture and feeds it an empty context. The thread's contribution to macro-F1 therefore lies between 0.005 and 0.033, and where it falls in that range depends on whether encoder capacity is removed along with the context. The 0.08 gap between CORAL and CORN suggests that the shared untuned configuration may disadvantage some heads.

\BfPara[app:errorconc]{Error concentration}
The body and \Cref{fig:entropy}(b) both use the severity-label definition of contested comments, those where no single severity level receives at least 4 of the 5 annotator labels. Because a cyberbullying comment carries at most 2 not-cyberbullying labels, any count of 4 or 5 must fall on one severity level, so this partition coincides exactly with the contested rows of \Cref{tab:dataset}. Under it, contested comments receive 85\% of the errors at an error rate 1.91 times the clear-comment rate.

Two looser definitions give larger numbers. A raw-ballot definition, which treats a comment as clear whenever its severity labels are unanimous and so moves 662 comments carrying non-CB labels onto the clear side, gives 2.05 times with 81\% of the errors. A purely entropy-based definition gives 2.3 times with 91\% of the errors. The severity-label figures the body reports are therefore the most conservative of the 3.

Both quantities behind the correlation are noisy, a 5-label entropy estimate on the human side and a narrow entropy range on the model side, so the attainable correlation is attenuated. We read $\rho \approx 0.12$ as weak and not as a precise bound.

\BfPara[app:deferral]{Deferral budget sweep}
We report accuracy on the kept comments because per-class F1 grows unstable as the kept set shrinks. At a 30\% budget the confidence of \ours{} gives 0.770 accuracy, human label entropy gives 0.750, and their combined rank reaches 0.772, which is indistinguishable from confidence alone (paired bootstrap CI $[-0.005, +0.009]$, and a session-level cluster bootstrap gives $[-0.006, +0.010]$). The same holds at smaller budgets.

When half of all comments are deferred, the combined rank reaches 0.828 against 0.797 for confidence alone. We read this larger-budget gain with care. Deferring high-entropy comments removes exactly the items whose majority labels are fragile, so part of the measured gain reflects the reliability of the evaluation label and not better routing of model errors.

\BfPara[app:capacity]{Calibration capacity}
Two reference points bound what the label-fitted map does. A 6-parameter lookup that outputs the class-conditional mean label distribution given the twin's predicted class reaches only 0.154 Brier, so the map is doing more than un-sharpening predictions toward class priors. A label-fitted 2-layer head of roughly a thousand parameters, used in place of the 12-parameter map, improves the twin only to 0.143, which is marginally past the 12-parameter map's 0.144. Added post-hoc capacity is therefore not the limitation.

Functional form does matter. A single temperature fitted to the label distribution instead of to the majority label reaches only 0.161, closing about a sixth of the gap between majority calibration at 0.164 and the full map at 0.144. Most of the label-specific gain therefore comes from the map's 12 parameters and not from a single rescaling toward the labels.

The majority-fitted and label-fitted Dirichlet maps of the decomposition are both fitted out of fold, which makes their comparison fair and distinguishes them from the in-fold majority-fitted maps that \Pref{app:calibmaps} uses only as a confidence-calibration baseline.

\BfPara[app:floor]{The sampling-noise floor}
A Brier score of 0.069 is the sampling-noise floor for this corpus. We estimate it 2 ways that agree: the plug-in per-comment multinomial variance $(1 - \sum_k \hat{q}_k^2)/n$ with $n = 5$ averaged over comments, and a simulation that resamples 5 labels from each comment's empirical distribution and scores the true distribution against the resample (0.069, 95\% CI $[0.067, 0.071]$).

This is a loose lower bound in 2 senses. It is the floor for a model that could predict each comment's latent label distribution, but a model that reads only the text may not be able to recover that latent distribution, so the attainable floor may be higher. The percentages of the closable gap we report are therefore fractions of a possibly generous upper bound on what is closable. Against a higher, text-conditional floor the same absolute gains close a larger fraction, so calibration's share falls and the small gain from label training grows in relative terms.

We report the sampling floor because it is the one quantity we can estimate without assuming a model class, and we treat the ordering of the steps, which is robust to the choice of floor, as the result rather than any single percentage. Read against this floor, the residual after label calibration is 0.075 in Brier. Training on the labels recovers 0.007 of that residual, about a tenth, and the remaining 9 tenths is not addressable by more calibration capacity or by training on the labels. That 0.007 is the same absolute gain the body highlights, about 3\% of the full closable gap of 0.225.

\BfPara[app:controls]{Control margins}
Under symmetric label calibration the trained-in controls give the same picture as the matched twin. Against calibrated label smoothing the margins are Brier $-0.006$ (CI $[-0.009, -0.004]$, 9 of 10 folds), log score $-0.007$ (CI $[-0.012, -0.001]$, 8 of 10), and TV $-0.003$ (CI $[-0.005, -0.001]$, 9 of 10). Against calibrated SORD they are Brier $-0.006$ (CI $[-0.008, -0.003]$, 9 of 10), log score $-0.006$ (CI $[-0.010, -0.002]$, 7 of 10), and TV $-0.003$ (CI $[-0.005, -0.001]$, 7 of 10). The full per-fold CIs for the matched-twin comparison are Brier $[-0.010, -0.005]$, log score $[-0.014, -0.004]$, and TV $[-0.008, -0.003]$.

Of the 9 comparisons (3 baselines by 3 metrics), only the matched-twin Brier comparison survives Holm correction at $\alpha = 0.05$ ($p = 0.009$). The remaining 8 have Holm-corrected $p$ values from 0.05 to 0.34 and are not individually significant, so the sign consistency, the held-out-label replication and the seed check carry the inferential weight.

\BfPara[app:severe]{The severe class}
On the severe class the 2 pipelines are close and the fold support is weak. The severe-class F1 of \ours{} before calibration is 0.675 against the twin's 0.663, but it leads in only 4 of 10 folds. On the 487 severe-majority comments the calibrated twin has the lower full Brier by about 0.006 in 6 of 10 folds. The severe class is therefore where the training signal is neither a clear help nor a clear harm, and we make no claim for it either way.

\BfPara[app:margins]{Held-out-label, seed, and LLM margins}
Scored against the single held-out label under symmetric label calibration, \ours{} remains below its retrained hard-label twin on log score ($-0.010$, CI $[-0.014, -0.006]$, 9 of 10 folds) and Brier score ($-0.009$, CI $[-0.011, -0.007]$, 10 of 10 folds). On the 1,757 no-plurality comments the log-score margin holds ($-0.024$, CI $[-0.039, -0.011]$, 8 of 10 folds), while the Brier interval crosses zero.

Both pipelines predict the held-out label better than the empirical distribution of the remaining labels does (Brier 0.49 against 0.61). A 2-to-4-label empirical distribution is a high-variance predictor of a single held-out label, so this baseline is weak by construction, but the comparison still shows that the models generalize beyond the labels they were given.

Retraining \ours{} and its twin under 5 training seeds across all 10 folds gives a mean calibrated Brier margin of $-0.0089$ (SD $0.0007$), negative at every seed, and a mean macro-F1 advantage of $+0.0167$ (SD $0.0016$), positive at every seed.

For the LLM references, zero-shot GPT-4o-mini reaches a mean confidence of 0.96 against label distributions it never saw, and Qwen3.6-27B is better but still far off (Brier 0.342). The 12-parameter label-fitted map moves GPT-4o-mini to 0.160 and Qwen to 0.149. \Ours{} with the same calibration stays below GPT-4o-mini on total variation and Brier in 10 of 10 folds (Brier $-0.022$, CI $[-0.026, -0.018]$). We omit log score for this comparison because GPT-4o-mini's log score is dominated by the top-15 clipping and is uninformative (\Pref{app:llm}). Against Qwen it stays below on Brier ($-0.011$, CI $[-0.017, -0.006]$, 9 of 10) and TV ($-0.011$, CI $[-0.014, -0.007]$, 9 of 10), with the 2 log scores statistically tied.

\BfPara[app:disagreement]{Disagreement characterization in full}
We divide the 9,553 CB comments into high- and low-disagreement groups by the top and bottom entropy terciles of their label distributions. Terciles give 2 clearly separated groups while the ambiguous middle third is dropped, and the topic regression uses entropy continuously over all comments, so the contrasts do not depend on this split.

High-disagreement comments appear earlier in their threads, with a median position of 36 against 50 for low-disagreement comments (Mann-Whitney \citep{mann1947test} $p < 0.0001$), where position is the raw sequence index and not a fraction of thread length. Early comments may offer less accumulated context about who is speaking and whether the exchange is escalating, though annotators worked within full sessions, so we read this as a hypothesis and not an established cause. High-disagreement comments are also slightly longer, with a median of 14 words against 12 ($p < 0.0001$), but the length association does not survive controls for topic and position and we read it as weak. Comments cluster within sessions while these tests treat them as independent, so the $p$ values are anticonservative and we read them as descriptive.

For the topic regression, gender ($+0.098$, $p<10^{-28}$), social status ($+0.070$) and political ($-0.148$, $p<10^{-12}$) survive the controls, while the sexual-content association is explained away by co-occurring factors. Because the comment-level tests ignore session clustering, we also resample at the session level. The gender ($+0.09$, 95\% cluster-bootstrap CI $[+0.07, +0.12]$), social-status ($+0.05$, CI $[+0.02, +0.08]$) and political ($-0.16$, CI $[-0.29, -0.02]$) entropy differences all hold. The patterns are also not artifacts of excluding the no-plurality comments. Recomputed over all 11,310 majority-CB comments including the 1,757 without a severity plurality, gender remains the highest-disagreement topic (0.894 against 0.769), political the lowest (0.664 against 0.810), the position gap persists (median 40 against 50, $p<0.0001$), and race-targeted content becomes elevated after correction.

Disagreement concentrates at the boundaries between adjacent levels, and 68.7\% of all disagreeing label pairs differ by one level. The direction of dissent is asymmetric. When the majority grades a comment mild, nearly all dissenting labels move one step to moderate (96\%) and almost none jump to severe. When the majority grades a comment severe, 27\% of dissenting labels fall all the way to mild. The severe class therefore holds 2 kinds of comments, those everyone recognizes and a divisive minority that some annotators read as harmless. Without annotator identities we cannot rule out scale-endpoint effects or occasional inattention behind these long jumps.

\input{figures/fig_reliability}

\input{figures/fig_forest}

\input{figures/fig_entropy}

\input{figures/fig_examples}

%% file: tables/tab_backbones.tex
\begin{table}[t]
\centering
\small
\setlength{\tabcolsep}{4pt}
\begin{tabular}{lcccc}
\toprule
Backbone & Macro-F1 & QWK & MAE & Off-by-1 \\
\midrule
DeBERTa-v3-base            & \textbf{0.652} & 0.730 & \textbf{0.240} & 0.970 \\
~~+ context ($K{=}5$)      & 0.650 & \textbf{0.734} & 0.244 & 0.971 \\
HateBERT                   & 0.648 & 0.726 & 0.245 & 0.972 \\
Twitter-RoBERTa            & 0.645 & 0.725 & 0.242 & 0.972 \\
RoBERTa-base               & 0.641 & 0.714 & 0.253 & 0.968 \\
ToxicBERT                  & 0.634 & 0.711 & 0.251 & 0.967 \\
BERT-base                  & 0.617 & 0.703 & 0.261 & 0.971 \\
\midrule
Pairwise+CORAL             & 0.587 & 0.716 & --    & --    \\
\bottomrule
\end{tabular}
\caption{Backbone comparison under one shared training configuration (fine-tuned flat classifier, 4-class scope, mean over the session-level folds). DeBERTa-v3-base is best on macro-F1, MAE and cross-fold variance and ties for best QWK. The prior-art Pairwise+CORAL ordinal competitor is shown for reference. The 5-comment context window changes QWK by $+0.005$ at higher variance. The 0.617--0.652 macro-F1 range previews the accuracy ceiling of \Pref{sec:saturation}.}
\label{tab:backbones}
\end{table}

%% file: figures/fig_reliability.tex
\begin{figure*}[t]
\centering
\includegraphics[width=0.95\textwidth]{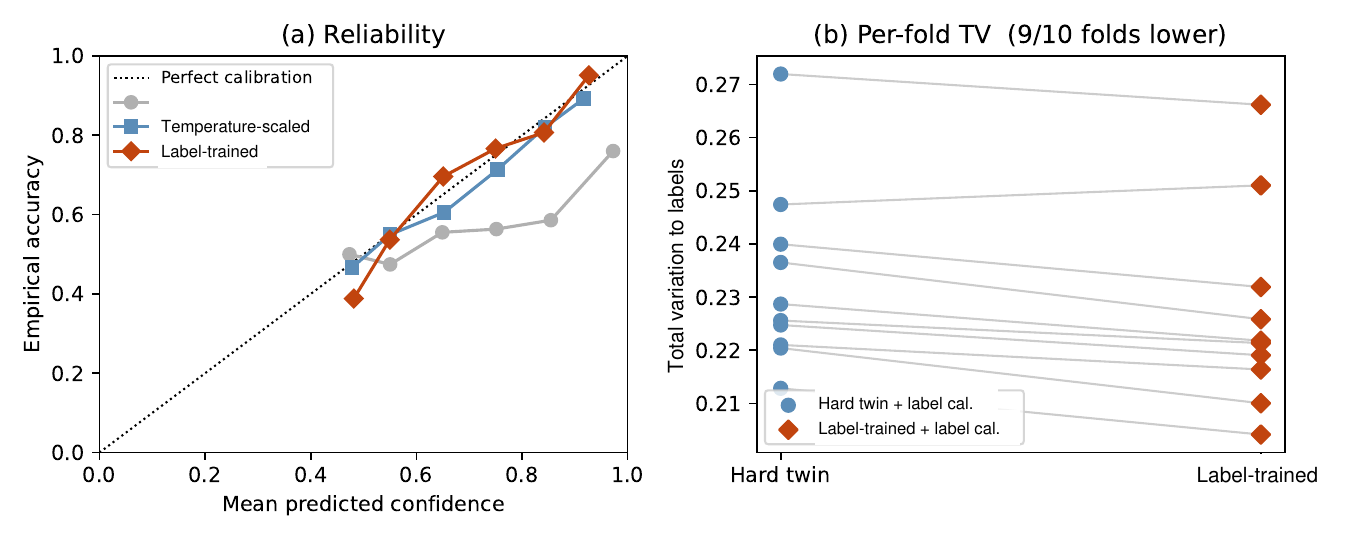}
\caption{(a) Reliability diagram. Gray circles (\mcirc{figgray}) show the flat 3-class baseline, blue squares (\msq{figblue}) its temperature-scaled version and orange diamonds (\mdia{figorange}) \ours{}. The dotted line is perfect calibration. The raw baseline sits far below the diagonal, so its confidence overstates its accuracy, while the scaled baseline and \ours{} both track the diagonal. (b) Total variation to the human label distribution per fold under the same label-fitted calibration for both systems, with lines connecting the same fold. \Ours{} (\mcirc{figorange}) is below its hard-label twin (\mcirc{figblue}) in 9 of 10 folds. The full comparison is in \Cref{tab:calibration} and \Cref{fig:forest}.}
\label{fig:reliability}
\end{figure*}

%% file: figures/fig_forest.tex
\begin{figure*}[t]
\centering
\includegraphics[width=0.95\textwidth]{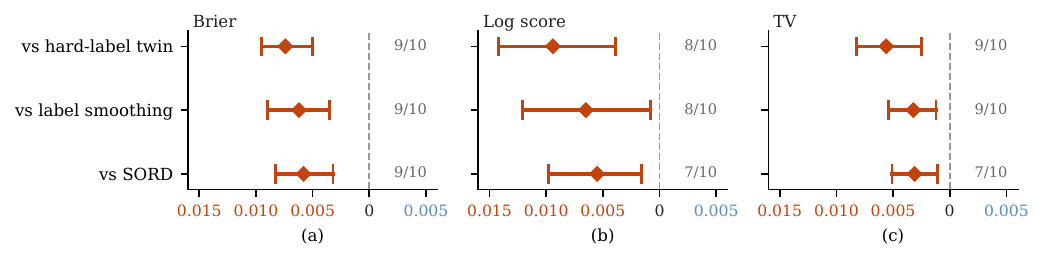}
\caption{Paired per-fold differences, \ours{} with label calibration minus each label-free baseline with the same calibration, on (a) Brier score, (b) log score and (c) total variation. Diamonds (\mdia{figorange}) are mean differences over 10 folds with bootstrap 95\% intervals. The dashed line marks zero and the annotation gives the number of folds in which \ours{} is lower. Axis magnitudes are unsigned, with \textcolor{figorange}{orange} tick labels marking the side where \ours{} is better and \textcolor{figblue}{blue} tick labels the side where the baseline is better. Every interval lies on the distribution-trained model's side.}
\label{fig:forest}
\end{figure*}

%% file: figures/fig_entropy.tex
\begin{figure*}[t]
\centering
\includegraphics[width=0.95\textwidth]{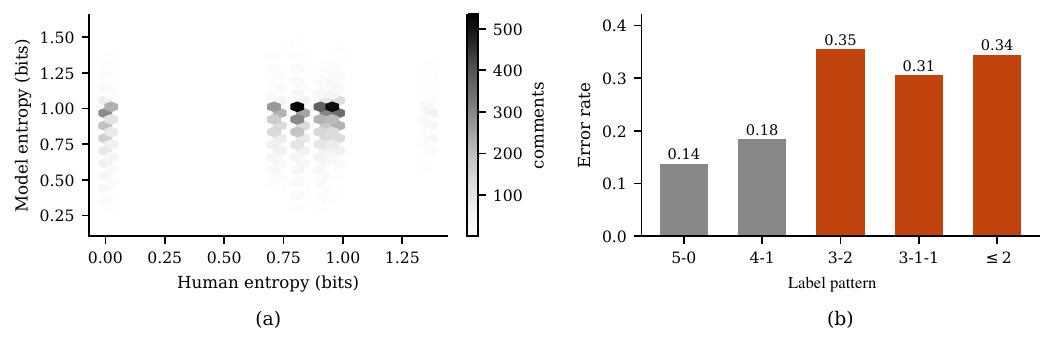}
\caption{(a) Model entropy against human label entropy for all 9,553 comments, where darker cells hold more comments. The 2 signals are nearly uncorrelated ($\rho = 0.12$). (b) Error rate by severity-label pattern. Gray bars (\swatch{figgray}) are clear patterns and orange bars (\swatch{figorange}) are contested patterns, those where no single severity level holds 4 of the 5 annotator labels, where the error rate is 1.9 times the clear-pattern rate.}
\label{fig:entropy}
\end{figure*}

%% file: figures/fig_examples.tex
\begin{figure*}[t]
\centering
\includegraphics[width=0.95\textwidth]{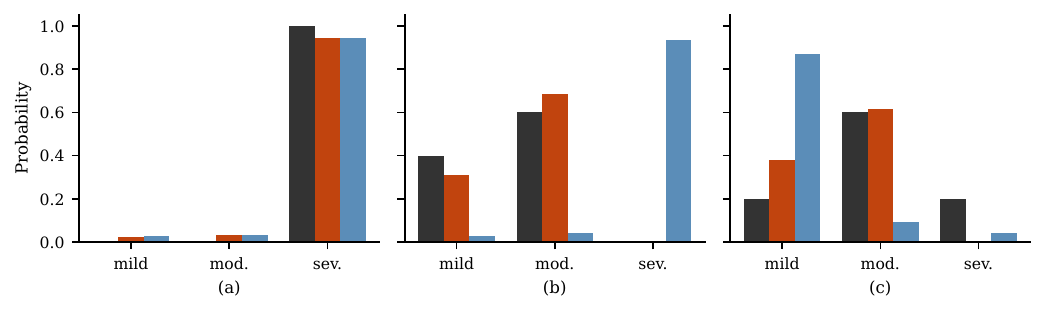}
\caption{Predicted distributions against the 5 human labels for 3 comments: (a) unanimously severe, ``@user you are really sh*tty and go die''; (b) split 2-3-0, ``tell that fool @user to lay off the haterade. Get a life...''; (c) split 1-3-1, ``A little f*ggish much @user''. Dark bars (\swatch{figdark}) show the human label distribution, orange bars (\swatch{figorange}) the prediction of \ours{} and blue bars (\swatch{figblue}) the temperature-scaled flat baseline. On the unanimous comment all 3 agree. On the split labels the scaled baseline concentrates its mass on a single class while \ours{} follows the shape of the human distribution.}
\label{fig:examples}
\end{figure*}